\documentclass[aps,prl,reprint,superscriptaddress,floatfix]{revtex4-2}
\usepackage{graphicx} 
\usepackage[utf8]{inputenc}

\usepackage[english]{babel}
\usepackage{ifthen}
\usepackage{xcolor}
\usepackage{amsmath}
\usepackage{mathtools}
\usepackage{braket}
\usepackage{siunitx}
\usepackage[mode=buildnew]{standalone}
\usepackage{multirow}

\usepackage[bookmarksopen]{hyperref}
\hypersetup{colorlinks,linkcolor=darkblue,citecolor=darkblue,urlcolor=darkblue}
\usepackage{upgreek}
\newboolean{ShowChanges}\setboolean{ShowChanges}{true}  
\provideboolean{ShowChanges}

\newcommand{\narrowtimes}{\medmuskip=1mu\times}
\newcommand{\SIangfreq} [2] {\ensuremath{2\pi\narrowtimes\SI{#1}{#2}}}
\DeclareSIUnit[]\bohrradius
{\text{\ensuremath{a_{0} }}}
\DeclareSIUnit[]\elemcharge
{\text{\ensuremath{e }}}

\newcommand{\FigRef}[1]{Fig.~\ref{#1}}

\newcommand{\SubFigRef}[2]{Fig.~\ref{#1}(#2)}

\newcommand{\SubFigRefCap}[2]{Figure~\ref{#1}(#2)}
\newcommand{\EqnRef}[1]{Eq.~(\ref{#1})}

\newcommand{\AffQVIL}[0]{Quantum Valley Ideas Laboratories, 485 Wes Graham Way, Waterloo, ON N2L 0A7, Canada}
\newcommand{\AffPI}[0]{5. Physikalisches Institut, Universität Stuttgart, Pfaffenwaldring 57, 70569 Stuttgart, Germany}

\newboolean{ShowComments}\setboolean{ShowComments}{true}  
\provideboolean{ShowComments}

\newcommand{\ShowMyCommnt}[1]{%
\ifShowComments%
#1%
\fi%
}
\newcommand{\NewComment}[2]{%
\expandafter\newcommand\csname#1Comment\endcsname[1]{\ShowMyCommnt{\textcolor{#2}{##1}}}%
}
\NewComment{Harald}{violet}
\NewComment{Jim}{blue}
\NewComment{Matze}{red}
\usepackage[english]{babel}
\definecolor{darkblue}{rgb}{ 0, 0, 0.7843137254901961}

\begin{document}
\title{All-optical radio-frequency phase detection for Rydberg atom sensors using oscillatory dynamics}
\author{Matthias Schmidt}
\affiliation{\AffQVIL}
\affiliation{\AffPI}
\author{Stephanie M. Bohaichuk}
\affiliation{\AffQVIL}
\author{Vijin Venu}
\affiliation{\AffQVIL}
\author{Ruoxi Wang}
\affiliation{\AffQVIL}
\author{Harald Kübler}
\affiliation{\AffQVIL}
\affiliation{\AffPI}
\author{James P. Shaffer}
\email{jshaffer@qvil.ca}
\affiliation{\AffQVIL}

\date{\today}


\begin{abstract}
Rydberg atom radio frequency sensors are a unique platform for precision electromagnetic field measurement, e.g. they have extraordinary carrier bandwidth spanning MHz-THz and can be self-calibrated. These photonic sensors use lasers to prepare and read out the atomic response to a radio frequency electromagnetic field. Most work on Rydberg atom sensors centers on radio frequency electric field strength because the sensor functions as a square law detector, unless an external radio frequency heterodyning field is used. A heterodyning field acts as a local oscillator and enables phase read out at the expense of the radio frequency equipment necessary to generate it. In order to overcome the disadvantages of a radio frequency local oscillator, we investigate all-optical phase-sensitive detection using a five-level closed-loop excitation scheme. We show that under finite detuning of the loop fields, the atomic response oscillates at the frequency of the detuning. 
The oscillation is transferred to a probe laser absorption signal. 
The phase, frequency and amplitude of the radio frequency signal are imprinted on the oscillatory dynamics and can be determined using demodulation and matched filter techniques applied to the probe laser transmission signal.
\end{abstract}
\maketitle

Rydberg atoms have emerged as a promising platform for radio frequency (RF) field sensing~\cite{Sedlacek2012,Holloway2014,Fan2014,Holloway2017,Adams_2019,Sedlacek2013}. Self-calibration, extraordinary carrier bandwidth and the construction of sensors of practically any shape and size using vapor cells that are electromagnetically transparent are some of their unique advantages when compared to conventional antenna systems~\cite{Fan_2015,Adams_2019, Noaman2023}. Important RF measurements such as polarization and angle of arrival of the RF wave have been investigated~\cite{Sedlacek2013, Simons2019, Holloway21AngleOfArrival, schlossberger2025angleofarrival}. Progress in pulsed detection has made Rydberg sensors interesting for applications using modulated RF fields~\cite{Meyer2018, Anderson2021, Jiao_2019, Holloway2021AMR, Song2019,Bohaichuk2022}. Decoding modulated fields generally requires phase sensitivity as many encoding schemes use the in-phase ($I$) and quadrature ($Q$) components of the signal. We describe a method for all-optical phase detection that can directly decode the $I$ and $Q$ components of an RF signal. 

All-optical phase detection can be realized through closed-loop excitation schemes in multilevel atomic systems~\cite{Morigi02, Morigi07PhaseVapor, RaithelPhase22, HollowayPhase23, borowka2025PhaseEp, kasza2024PhaseTheory}. We theoretically investigate the oscillatory dynamics of the atomic response that occurs when the excitation loop is operated under a finite detuning of at least one of the loop coupling fields. Under these conditions, the oscillations are transferred to a probe laser transmission signal. The oscillations of the probe laser transmission, have not, to our knowledge, been described in detail and can be used to determine the phase, amplitude and detuning of the target RF field. Demodulation yields $I$ and $Q$ directly. A closed form formula within the weak probe and strong coupling approximations is given and describes the behavior well. 

Rydberg atom RF field sensing relies on measuring spectroscopic changes induced by an incident RF field. The transmission changes of a probe laser beam passing through an atomic vapor are used to derive the properties of the RF field~\cite{Sedlacek2012,Holloway2014,Fan2014,Holloway2017,Adams_2019,Sedlacek2013, Bohaichuk2022, Bohaichuk2023threephoton, Venu2025}. 
All-optical approaches determine the RF field amplitude through Autler-Townes (AT) splitting~\cite{Sedlacek2012, Holloway2017, Gordon2014} or probe laser transmission amplitude-based techniques~\cite{Schmidt2024, Bohaichuk2023threephoton,Venu2025,Kuebler2018}. All-optical sensitivities of ${240}\,\textrm{nV}\,\textrm{cm}^{-1}\,\textrm{Hz}^{-1/2}$ have been realized with continuous and pulsed RF fields in a two-photon scheme~\cite{Kumar2016, Kumar2017, Bohaichuk2022, Sapiro_2020} and ${45}\,\textrm{nV}\,\textrm{cm}^{-1}\,\mathrm{Hz}^{-1/2}$ has been achieved using a three-photon scheme~\cite{Venu2025}. Heterodyne detection techniques, where an auxiliary RF field is used along with the target RF field, have been established and have demonstrated higher sensitivities, $\SI{12}\,\textrm{nV}\,\textrm{cm}^{-1}\,\textrm{Hz}^{-1/2}$~\cite{Cai2022}. A disadvantage of heterodyne methods, when compared to all-optical measurement, is the necessity for conventional RF equipment to supply a local oscillator over $100\,$GHz range in order to take advantage of the broad carrier bandwidth of the sensor. A notable advantage of the heterodyne technique is that it supports phase-sensitive measurements~\cite{Bang2022, Cai2022, Jing2020, Simons2019, Simons2021, Cui2023, Anderson2019RydbergAF, Adams_2019}. All-optical phase detection enables the Rydberg system to be constructed without an RF heterodyning field. Our work advances all-optical sensing so that Rydberg sensors can achieve the $>100\,$GHz carrier bandwidth enabled by the atom, while simultaneously measuring both RF phase and amplitude.

In the optical interferometry approach described here, we use an “optical + RF signal under test” closed loop, with one atomic energy level of the loop coupled to a probe transition. The transitions around the loop add in clockwise and counter-clockwise paths, starting from the upper level of the probe laser transition. When adding the two loop pathway amplitudes, the phase contributions are equal and opposite signed, while the amplitudes are the same, resulting in a sinusoidal behavior dependent on the phase accumulated in each loop. When one of the coupling fields making up the loop is detuned, the phase becomes time dependent and oscillates at the frequency of the detuning. The time dependent phase directly affects the atomic response imprinted on the probe laser transmission. The oscillatory dynamics do not depend on any atomic decay rates and, thereby, are long-lived compared to any timescales set by the atomic medium. The oscillations are limited by the coherence of the driving fields.


\begin{figure*}
\centering
\includegraphics[width=.25\textwidth]{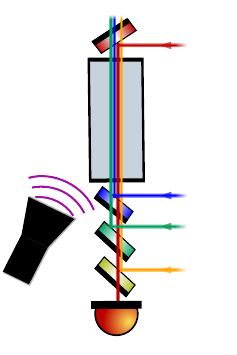} 
\includegraphics[width=.27\textwidth]{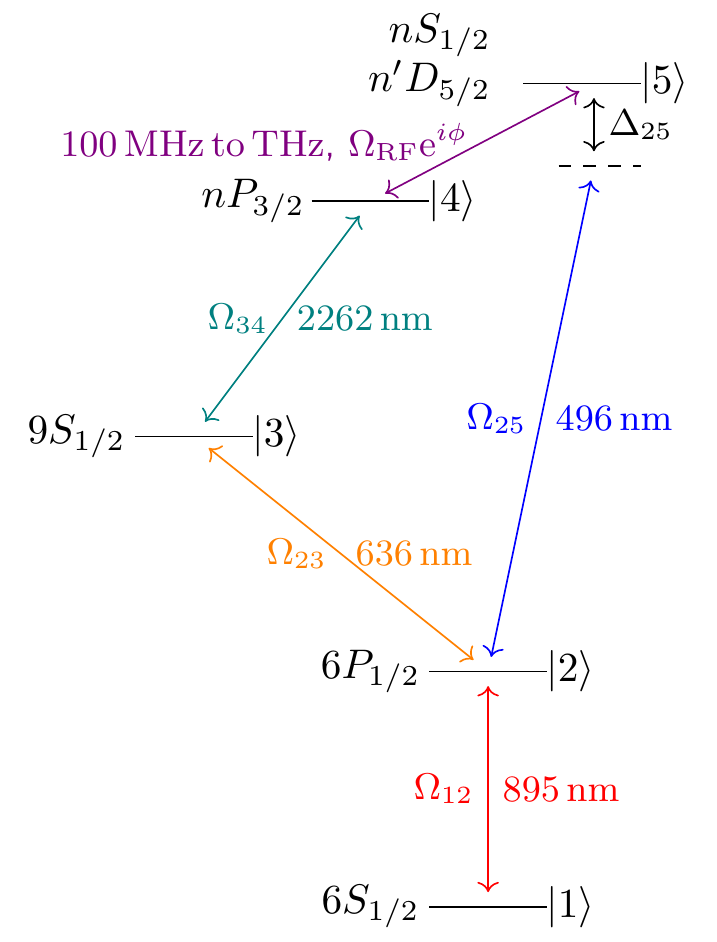} 
\includegraphics[width=.45\textwidth]{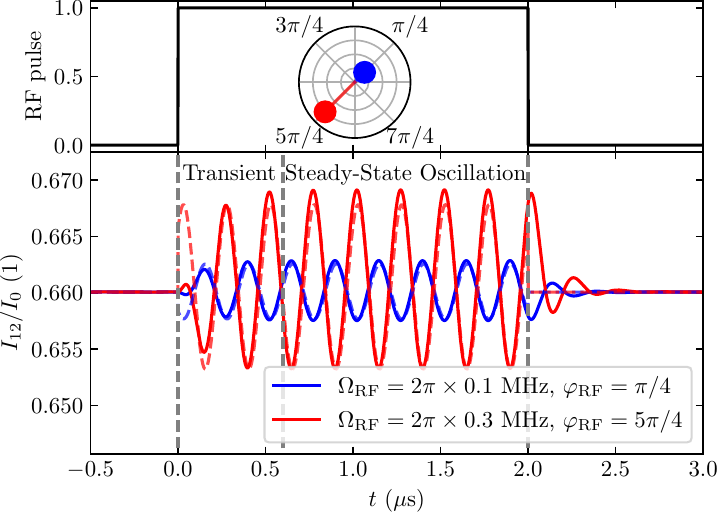}
\put(-500,165){(a)} \put(-375,165){(b)}\put(-230,165){(c)}
\put(-410,170){\textcolor{red}{$\Omega_{12}$}}
\put(-410,50){\textcolor{orange}{$\Omega_{23}$}}
\put(-410,72){\textcolor{teal}{$\Omega_{34}$}}
\put(-410,90){\textcolor{blue}{$\Omega_{25}$}}
\put(-500,100){\textcolor{violet}{$\Omega_\mathrm{RF}\mathrm{e}^{i\phi}$}}
\put(-505,30){RF antenna}
\put(-495,130){Vapor cell}
\put(-470,02){Photodetector}
\caption{(a) Setup. All lasers and the RF field are in a co-linear geometry. Thermal Cs atoms are contained in a vapor cell of length $L = \SI{3}{\centi \meter}$. The probe laser counter-propagates the other lasers to minimize Doppler mismatch. The intensity of the probe laser is measured by a photodetector after passing through the vapor cell.  
(b) Closed-loop excitation scheme. The closed loop consists of four optical laser fields and the RF field of interest. The probe laser transition is not part of the loop. 
For a finite blue laser detuning, $\Delta_{25}$, the probe intensity oscillates as shown in (c). (c) Upper figure: Sequence of two different RF pulses with different $\Omega_\mathrm{RF}$ and $\varphi_\mathrm{RF}$. The pulse envelope is shown in black. Lower figure: The probe laser transmission intensity dynamics oscillates at $\Delta_{25}/2\pi= \SI{4}{\mega \hertz}$. The dashed lines show the results from \EqnRef{Eq:WeakProbeExpression}. $\Omega_\mathrm{12}=\SIangfreq{100}{\kilo \hertz}$ and $\Omega_{23}=\Omega_{34}=\Omega_{25} = \SIangfreq{3}{\mega \hertz}$.}
\label{Fig1}
\end{figure*}
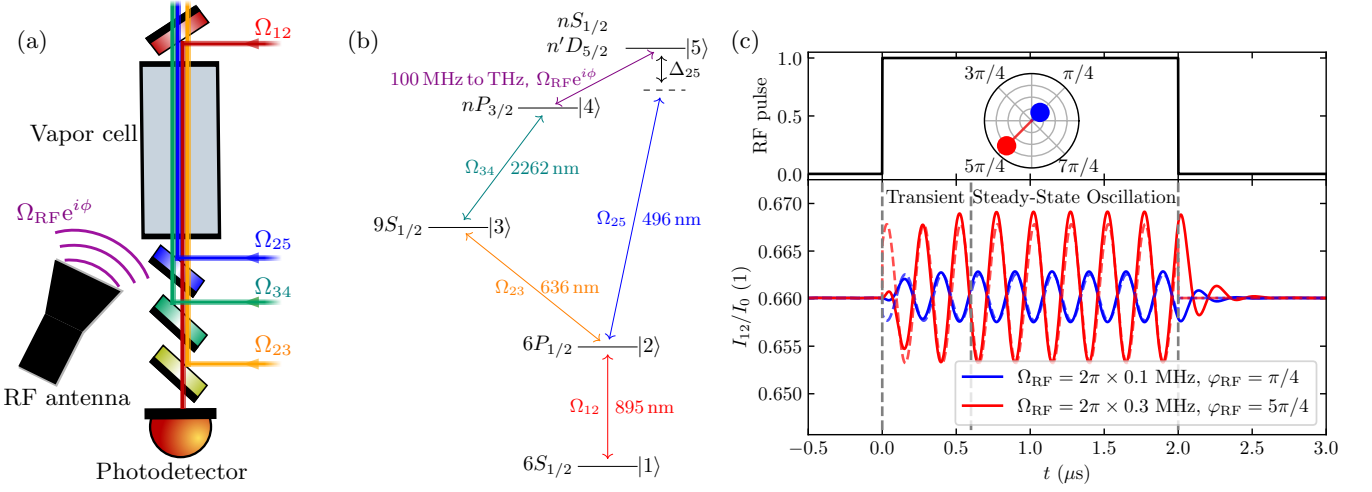

An example experimental setup is shown in \SubFigRef{Fig1}{a}. All lasers and the RF field are arranged in a co-linear geometry passing through a cesium vapor cell at room temperature. The probe laser counter-propagates the other lasers and the probe transmission is measured on a photodetector. More details on the laser geometry can be found in~\cite{supp}.

The excitation scheme consists of five levels driven by four laser fields and a target RF field that couples two Rydberg states,~\SubFigRef{Fig1}{b}. The probe laser is not part of the closed loop. In this work, the blue laser has a finite detuning, $\Delta_{25}$, leading to time-dependent dynamics that are utilized to detect the RF phase. All other lasers and the RF wave are considered to be on resonance. Other fields in the loop, or combinations thereof, can be used to induce the oscillatory phenomena.

To investigate the probe laser transmission, we numerically solve the Lindblad master equation
\begin{equation}
\dot{\rho}= \frac{i}{\hbar} [{\cal{H}}, \rho]-\cal{L},
\label{Eq:Theory_DensityMatrixEquation}
\end{equation}
where $\rho$ is the density matrix. Decay and dephasing are included in the Lindblad-operator, $\cal{L}$. For the example calculations, $\gamma_{2} = \SIangfreq{5}{\mega \hertz}$ for $6P_{1/2}$ and $\gamma_3 = \SIangfreq{1}{\mega \hertz}$ for $9S_{1/2}$. For the Rydberg states, we assume $\gamma_4 = \gamma_5 = \SIangfreq{20}{\kilo \hertz}$. The values for $\gamma_4$ and $\gamma_5$ are representative. We include transit time broadening of $\SIangfreq{200}{\kilo \hertz}$. 
The Hamiltonian, $\cal{H}$, models the five level closed-loop excitation scheme presented in \SubFigRef{Fig1}{b}. We apply the rotating wave approximation and choose the reference frame, without loss of generality, so that all phases of the driving fields are projected onto the RF transition~\cite{Morigi02, Morigi07PhaseVapor, supp}.
The detunings of the respective transitions are denoted $\Delta_{ij}$, while
the Rabi frequencies are denoted $\Omega_{ij} = \mu_{ij}E_{ij}/\hbar$, with the dipole moments, $\mu_{ij}$ and the E-field amplitudes $E_{ij}$. $i$ and $j$ indicate the corresponding transition, i.e. 1,..., 5. We use RF as a label for the Rydberg state transition to emphasize the field of interest, i.e. $\Delta_{45}:=\Delta_\mathrm{RF}$. The complex phase on the RF transition, $\phi$, simplifies to
\begin{align}
    \phi &= \Delta_{25}\cdot t + \varphi_\mathrm{RF}+ k_\mathrm{eff}\cdot z + k_\mathrm{eff}\cdot v \cdot t,
    \label{Eq:Phase}
\end{align}
under the assumption that only the blue laser has a finite detuning.
If the lasers all have zero phase and are phase locked relative to a reference such as a clock, $\varphi_{ij}=0$, then the phase shift is that of the RF target field. The phases of the lasers can be set to the reference phase by using a phase shifter such as a delay line or an electro-optic phase modulator. Details can be found in the Supplementary Materials~\cite{supp}.

$\phi$ is spatially independent in the co-linear geometry, $k_\mathrm{eff}=k_{23}+k_{34}+k_\mathrm{RF}-k_{25}=0$. The angle of arrival of the RF field to the optical axis should be small enough so that the projection of $k_\mathrm{RF}$ on the optical axis leads to a spatial periodicity that is larger than the length of the vapor cell. The probe laser counter-propagates the other lasers for Doppler compensation.
The detunings are modified by the Doppler shifts where $\Delta_{ij} =\vec{k}_{ij} \cdot \vec{v}$. 
$|\vec{k}_{ij}|=2\pi/\lambda_{ij}$ denotes the magnitude of the wave vectors for the transition with wavelength $\lambda_{ij}$. Since the lasers are co-linear in our geometry, $\Delta_{ij}$ is a scalar. For a different laser geometry, where $k_\mathrm{eff}\neq0$, the atoms belonging to different velocity classes have different Doppler shifts, so their complex phase oscillates at different frequencies, \EqnRef{Eq:Phase}. 

The complex atomic response, $\epsilon$, is calculated by integrating over all velocities weighted by the Boltzmann distribution, $P(v)$, at temperature, $T$,
\begin{equation}
\epsilon =\eta + i \alpha =  A k_\mathrm{12} \frac{\gamma_2}{\Omega_\mathrm{12}} \int_{-\infty}^\infty\rho_{12}(v)P(v) \, \mathrm{d}v,
\label{Eq:AtomicResponse}
\end{equation}
where $A=3 n\lambda^3_\mathrm{12}/16\pi^2$ is a dimensionless factor consisting of the atomic density, $n$, and the probe transition wavelength, $\lambda_\mathrm{12}$. $\eta$ and $\alpha$ correspond to the refractive index and absorption coefficient, respectively.
The density matrix element $\rho_{12}$ describes the coherence of the probe transition.
The intensity of the probe laser is expressed as $I_{12}(t) = I_0\, \mathrm{e}^{-\alpha(t) L}$, with $I_0 = c\left|E_{12}\right|^2/8\pi$.

\SubFigRefCap{Fig1}{c} shows the time evolution of the probe laser intensity, $I_{12}(t)$, for two different $\Omega_\mathrm{RF}$ and $\varphi_\mathrm{RF}$ with $\Delta_{25} = \SIangfreq{4}{\mega \hertz}$~\cite{supp}. In the upper panel, a normalized $\SI{2}{\micro \second}$ RF pulse is shown. The inset of the polar plot shows the relative phase and amplitude of the signals. The time evolution (lower panel) shows the oscillatory dynamics starting at $t= 0$. The oscillation frequency is $\SI{4}{\mega \hertz} =\Delta_{25}/2\pi$. If the loop detuning is known, the RF detuning frequency can be derived from the oscillation frequency. The $\pi$ phase shift between the two RF pulses is observed in the relative phase of the $I_{12}(t)$ oscillations. 

During the pulse, we distinguish two regions(marked by the dashed lines in \SubFigRef{Fig1}{c}); a transient period and a steady-state oscillation period. At the beginning of the pulse, the transient period, there are two components to the evolution. The atom starts the oscillatory dynamics associated with the interferometric loops, but also responds to the change in the RF field vector on the Rydberg state transition. The population distribution of the two RF dressed states relaxes as the phase change rotates the RF field vector in the $u-v$ plane of the Bloch sphere, causing the Bloch vector to relax in response to new dressed states. The decay of the transient is determined by the various decay and dephasing mechanisms, such as transit time, spontaneous emission and collisions~\cite{Bohaichuk2022}. The temporal decay of the transients ultimately limits the bandwidth of the sensor and is set by the atomic decay rates and Rabi frequencies~\cite{supp}. The transient can be distinguished from the loop phase change, leading to the oscillatory behavior, in that the relative contributions of the two Rydberg states to the dressed states changes in time. 
After the transients decay, $I_{12}(t)$ oscillates with maximum amplitude until the RF pulse ends. When the pulse ends, $I_{12}(t)$  decays to the steady-state transmission. 

The oscillatory dynamics are not symmetric around the steady-state transmission, but shift towards increased probe transmission. The shift is due to absorptive loop processes that also affect the probe laser absorption. The shift is strongly reduced in a thermal system due to Doppler averaging, but is more dominant in a zero temperature calculation~\cite{supp}.

\begin{figure*}
\includegraphics[width = \textwidth]{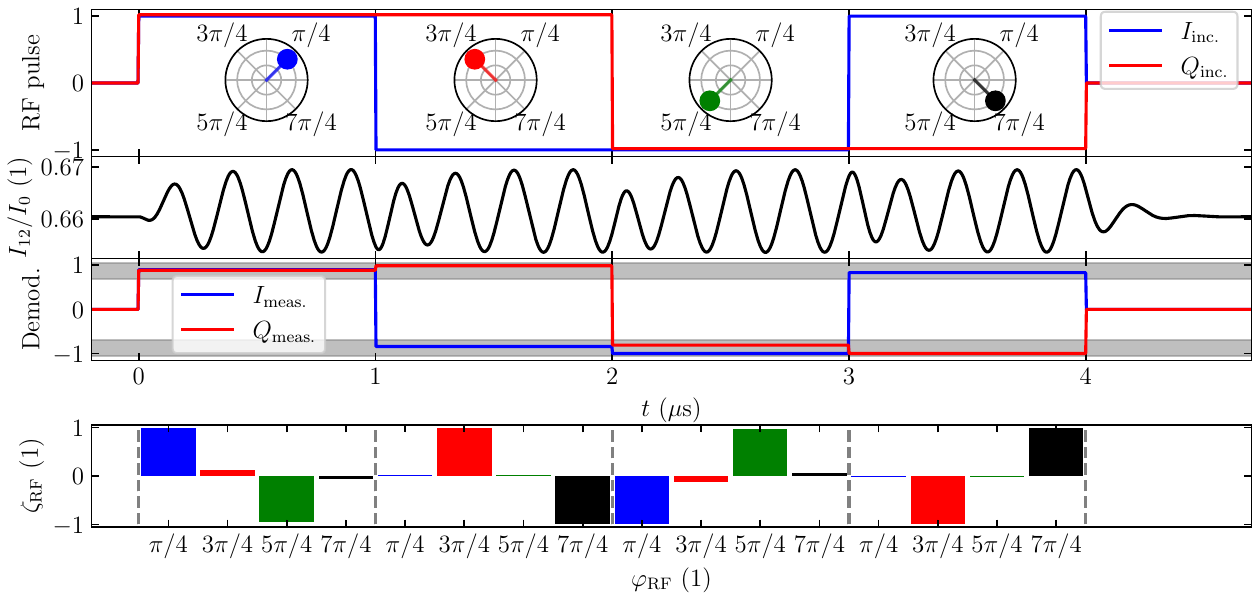}
\put(-520,235){(a)}\put(-520,180){(b)}\put(-520,130){(c)}\put(-520,70){(d)}
\vspace{-0.3cm}
\caption{(a) $\SI{4}{\micro \second}$ RF pulse with $\SI{1}{\micro \second}$ chip width. $\Omega_\mathrm{RF} = \SIangfreq{0.3}{\mega \hertz}$. (b) Probe intensity for the corresponding RF pulse in (a). (c) Change in phase for each chip in (b) evaluated using \EqnRef{Eq:IQ}. The small changes in the demodulated signal, e.g. for $Q_\mathrm{meas.}$ at $\SI{1}{\micro \second}$, are due to the transient dynamics. (d) Matched filter decoding of the signal in (b). $\Delta_{25}/2\pi= \SI{4}{\mega \hertz}$, $\Omega_\mathrm{12}=\SIangfreq{100}{\kilo \hertz}$ and $\Omega_{23}=\Omega_{34}=\Omega_{25} = \SIangfreq{3}{\mega \hertz}$.}
\label{Fig2}
\end{figure*}

The weak probe approximation can be applied to the Lindblad equation to obtain a closed form expression for the dynamics. 
In the weak probe limit, $\Omega_\mathrm{12} \ll \Omega_{ij}$ for $ij\neq12$~\cite{supp}, the atomic response shows the highest change for the incoming $\Omega_\mathrm{RF}$~\cite{Fleischhauer2005,Schmidt2024,Firstenberg2016, GeaBanacloche1995, Banacloche952, supp}. $\rho_{12}$ then depends linearly on $\Omega_\mathrm{12}$. Under these approximations,
\begin{widetext}
\begin{align}
    \rho_{12} =\dfrac{C_\mathrm{12}}{1+\dfrac{C_\mathrm{23}C_\mathrm{32} }{1+\dfrac{C_\mathrm{34}C_\mathrm{43} }{1+ C_\mathrm{RF}C^\prime_\mathrm{RF} }} + \dfrac{C_\mathrm{25}C_\mathrm{52}}{1+\dfrac{C_\mathrm{RF}C^\prime_\mathrm{RF}}{1+ C_\mathrm{34}C_\mathrm{43} }} - \dfrac{2C_\mathrm{23}C_\mathrm{34}C_\mathrm{RF}C_\mathrm{52} \mathrm{cos}(\phi) }{1+ C_\mathrm{34}C_\mathrm{43} + C_\mathrm{RF}C^\prime_\mathrm{RF} } }.
    \label{Eq:WeakProbeExpression}
\end{align}
\end{widetext}
In \EqnRef{Eq:WeakProbeExpression}, the coherences, $C_{jk} := i\Omega_{jk}/(\gamma_k+2i\Delta_{k})$ for $k>j$ and $i\Omega_{jk}=(i\Omega_{kj})^\star$~\cite{supp}.
In the definition of the coherences, the multi-photon detunings, $\Delta_{k}$, are considered.
The first two terms in the denominator of \EqnRef{Eq:WeakProbeExpression} correspond to the absorption of the left and right ladder pathways. The absorptive terms depend on $\Omega_\mathrm{RF}$ and explain why the oscillatory behavior is not symmetric around the steady-state transmission. The third term in the denominator is a superposition of the counter- and clockwise loops and is the only term explicitly phase dependent.
The amplitude of the dynamics depends on $\Omega_\mathrm{RF}$ and can be extracted if the other $\Omega_{ij}$ are known. Similar to \EqnRef{Eq:AtomicResponse}, we integrate \EqnRef{Eq:WeakProbeExpression} over the velocity to obtain the response of the atomic gas. More details on how \EqnRef{Eq:WeakProbeExpression} is derived can be found in~\cite{supp}.

The probe laser transmission calculated using \EqnRef{Eq:WeakProbeExpression} is shown in \SubFigRef{Fig1}{c} as dashed lines. The analytical expression matches the full numerics well, $<3\%$ deviation, over the steady-state oscillations in \SubFigRef{Fig1}{c}. The features that are not captured by \EqnRef{Eq:WeakProbeExpression} are the transients at the beginning and end of the RF pulse.

The oscillatory dynamics of $I_{12}(t)$ can be used to decode quadrature amplitude modulated (QAM) signals. We consider an incoming $\SI{4}{\micro \second}$ RF signal with a chip width of $\SI{1}{\micro \second}$ consisting of four different phases, $\varphi_\mathrm{RF} =  \{\pi/4, 3\pi/4, 5\pi/4, 7\pi/4\}$, corresponding to four symbols. Equivalently, the symbols can be expressed as $\Omega_\mathrm{RF} = |\Omega_\mathrm{RF}|/\sqrt{2} \left( I_\mathrm{inc.} + iQ_\mathrm{inc.} \right)$ with $\{I_\mathrm{inc.},Q_\mathrm{inc.}\}= \{\{1,1\},\{-1,1\},\{-1,-1\},\{1,-1\}\}$. We demonstrate that the signals can be decoded using demodulation at the detuning frequency and matched-filtering. 

\SubFigRefCap{Fig2}{a} shows an RF pulse sequence. The phase of the pulses switch instantaneously. $I_{12}(t)$ is shown in \SubFigRef{Fig2}{b}. During the phase change, the oscillatory dynamics of $I_{12}(t)$ adjusts at the time scale of the transient.
To demodulate the signal, we calculate,
\begin{align}
    I_\mathrm{meas.} &= \frac{1}{\tau} \int_0^\tau I_{12}(t)\,\mathrm{sin}(\Delta_{25} t + \vartheta)\,\mathrm{d}t \nonumber \\
    Q_\mathrm{meas.} &= \frac{1}{\tau} \int_0^\tau I_{12}(t)\,\mathrm{cos}(\Delta_{25} t + \vartheta)\, \mathrm{d}t.
    \label{Eq:IQ}
\end{align}
In \EqnRef{Eq:IQ} we add an offset phase, $\vartheta\approx 10^\circ$, to compensate the transient phase shift. The calculation reproduces the incoming sequence of the RF pulse, \SubFigRef{Fig2}{c}. The small changes in the demodulation signal, e.g. at $\SI{1}{\micro \second}$ for the $Q_\mathrm{meas.}$ curve, are associated with the transient behavior. It is important to note that the demodulation is performed on the probe transmission intensity which can be acquired on a square law detector, digitized and then demodulated in digital electronics.

The matched filter correlates the incoming signal with individual templates via
\begin{equation}
    \zeta_\mathrm{RF} = \frac{1}{\tau}\int_0^\tau f(t)\,I_{12}(t) \,\mathrm{d}t.
    \label{Eq:MatchedFilter}
\end{equation}
$f(t)$ denotes the matched filter template, that is calculated as a $\SI{1}{\micro \second}$ RF pulse with the respective phase and $\tau$ is the pulse length. The matched filter correlation output, $\zeta_\mathrm{RF}$, returns a value between $1$, for perfect correlation, and $-1$, for anti-correlation. The evaluation of \EqnRef{Eq:MatchedFilter} for each chip is shown in \SubFigRef{Fig2}{d}. The anti-correlation is explained by the $\pi$ phase shift between the respective phases of the chips. Pulses with a phase shift $\pm \pi/2$ return a correlation around zero indicating that the $I$ and $Q$ components can be separately read out. The nonzero contributions in the correlation plot are due to the transient response. 



The all-optical phase-sensitive RF field detection method presented here utilizes the oscillatory dynamics of a closed-loop excitation scheme under finite detuning. The oscillatory behavior can significantly advance Rydberg atom sensing because it enables simultaneous $I$ and $Q$ readout without an RF local oscillator. The detuning of the RF field can be determined using the same method and is compatible with our high sensitivity all-optical three-photon scheme \cite{Bohaichuk2023threephoton,Venu2025}. As the method detects the amplitude and phase of the RF via the probe intensity, the achievable sensitivities are similar to non-looped systems~\cite{Bohaichuk2023threephoton, Venu2025, Schmidt2024}. The phase of the RF field relative to the lasers imprints itself on the oscillatory dynamics of the atomic response and is robust to Doppler averaging. The phase detection of the GHz-THz RF wave gets frequency down-converted to the detuning, typically in the MHz range. We showed the successful reconstruction of an incoming QAM pulse sequence by demodulating and by applying a matched filter to $I_{12}(t)$ which is performed after readout on a photodetector.



We acknowledge funding from the Defense Advanced Research Projects Agency under HR001120S006 and the National Research Council Internet of Things: Quantum Sensors Challenge program through contract No. QSP-105-1.

\bibliography{biblio}

\begin{thebibliography}{43}%
\makeatletter
\providecommand \@ifxundefined [1]{%
 \@ifx{#1\undefined}
}%
\providecommand \@ifnum [1]{%
 \ifnum #1\expandafter \@firstoftwo
 \else \expandafter \@secondoftwo
 \fi
}%
\providecommand \@ifx [1]{%
 \ifx #1\expandafter \@firstoftwo
 \else \expandafter \@secondoftwo
 \fi
}%
\providecommand \natexlab [1]{#1}%
\providecommand \enquote  [1]{``#1''}%
\providecommand \bibnamefont  [1]{#1}%
\providecommand \bibfnamefont [1]{#1}%
\providecommand \citenamefont [1]{#1}%
\providecommand \href@noop [0]{\@secondoftwo}%
\providecommand \href [0]{\begingroup \@sanitize@url \@href}%
\providecommand \@href[1]{\@@startlink{#1}\@@href}%
\providecommand \@@href[1]{\endgroup#1\@@endlink}%
\providecommand \@sanitize@url [0]{\catcode `\\12\catcode `\$12\catcode `\&12\catcode `\#12\catcode `\^12\catcode `\_12\catcode `\%12\relax}%
\providecommand \@@startlink[1]{}%
\providecommand \@@endlink[0]{}%
\providecommand \url  [0]{\begingroup\@sanitize@url \@url }%
\providecommand \@url [1]{\endgroup\@href {#1}{\urlprefix }}%
\providecommand \urlprefix  [0]{URL }%
\providecommand \Eprint [0]{\href }%
\providecommand \doibase [0]{https://doi.org/}%
\providecommand \selectlanguage [0]{\@gobble}%
\providecommand \bibinfo  [0]{\@secondoftwo}%
\providecommand \bibfield  [0]{\@secondoftwo}%
\providecommand \translation [1]{[#1]}%
\providecommand \BibitemOpen [0]{}%
\providecommand \bibitemStop [0]{}%
\providecommand \bibitemNoStop [0]{.\EOS\space}%
\providecommand \EOS [0]{\spacefactor3000\relax}%
\providecommand \BibitemShut  [1]{\csname bibitem#1\endcsname}%
\let\auto@bib@innerbib\@empty
\bibitem [{\citenamefont {Sedlacek}\ \emph {et~al.}(2012)\citenamefont {Sedlacek}, \citenamefont {Schwettmann}, \citenamefont {K{\"u}bler}, \citenamefont {L{\"o}w}, \citenamefont {Pfau},\ and\ \citenamefont {Shaffer}}]{Sedlacek2012}%
  \BibitemOpen
  \bibfield  {author} {\bibinfo {author} {\bibfnamefont {J.~A.}\ \bibnamefont {Sedlacek}}, \bibinfo {author} {\bibfnamefont {A.}~\bibnamefont {Schwettmann}}, \bibinfo {author} {\bibfnamefont {H.}~\bibnamefont {K{\"u}bler}}, \bibinfo {author} {\bibfnamefont {R.}~\bibnamefont {L{\"o}w}}, \bibinfo {author} {\bibfnamefont {T.}~\bibnamefont {Pfau}},\ and\ \bibinfo {author} {\bibfnamefont {J.~P.}\ \bibnamefont {Shaffer}},\ }\bibfield  {title} {\bibinfo {title} {Microwave electrometry with rydberg atoms in a vapour cell using bright atomic resonances},\ }\href {https://doi.org/10.1038/nphys2423} {\bibfield  {journal} {\bibinfo  {journal} {Nature Physics}\ }\textbf {\bibinfo {volume} {8}},\ \bibinfo {pages} {819} (\bibinfo {year} {2012})}\BibitemShut {NoStop}%
\bibitem [{\citenamefont {Holloway}\ \emph {et~al.}(2014)\citenamefont {Holloway}, \citenamefont {Gordon}, \citenamefont {Jefferts}, \citenamefont {Schwarzkopf}, \citenamefont {Anderson}, \citenamefont {Miller}, \citenamefont {Thaicharoen},\ and\ \citenamefont {Raithel}}]{Holloway2014}%
  \BibitemOpen
  \bibfield  {author} {\bibinfo {author} {\bibfnamefont {C.~L.}\ \bibnamefont {Holloway}}, \bibinfo {author} {\bibfnamefont {J.~A.}\ \bibnamefont {Gordon}}, \bibinfo {author} {\bibfnamefont {S.}~\bibnamefont {Jefferts}}, \bibinfo {author} {\bibfnamefont {A.}~\bibnamefont {Schwarzkopf}}, \bibinfo {author} {\bibfnamefont {D.~A.}\ \bibnamefont {Anderson}}, \bibinfo {author} {\bibfnamefont {S.~A.}\ \bibnamefont {Miller}}, \bibinfo {author} {\bibfnamefont {N.}~\bibnamefont {Thaicharoen}},\ and\ \bibinfo {author} {\bibfnamefont {G.}~\bibnamefont {Raithel}},\ }\bibfield  {title} {\bibinfo {title} {Broadband rydberg atom-based electric-field probe for si-traceable, self-calibrated measurements},\ }\href {https://doi.org/10.1109/TAP.2014.2360208} {\bibfield  {journal} {\bibinfo  {journal} {IEEE Transactions on Antennas and Propagation}\ }\textbf {\bibinfo {volume} {62}},\ \bibinfo {pages} {6169} (\bibinfo {year} {2014})}\BibitemShut {NoStop}%
\bibitem [{\citenamefont {Fan}\ \emph {et~al.}(2014)\citenamefont {Fan}, \citenamefont {Kumar}, \citenamefont {Daschner}, \citenamefont {K\"{u}bler},\ and\ \citenamefont {Shaffer}}]{Fan2014}%
  \BibitemOpen
  \bibfield  {author} {\bibinfo {author} {\bibfnamefont {H.~Q.}\ \bibnamefont {Fan}}, \bibinfo {author} {\bibfnamefont {S.}~\bibnamefont {Kumar}}, \bibinfo {author} {\bibfnamefont {R.}~\bibnamefont {Daschner}}, \bibinfo {author} {\bibfnamefont {H.}~\bibnamefont {K\"{u}bler}},\ and\ \bibinfo {author} {\bibfnamefont {J.~P.}\ \bibnamefont {Shaffer}},\ }\bibfield  {title} {\bibinfo {title} {Subwavelength microwave electric-field imaging using rydberg atoms inside atomic vapor cells},\ }\href {https://doi.org/10.1364/OL.39.003030} {\bibfield  {journal} {\bibinfo  {journal} {Opt. Lett.}\ }\textbf {\bibinfo {volume} {39}},\ \bibinfo {pages} {3030} (\bibinfo {year} {2014})}\BibitemShut {NoStop}%
\bibitem [{\citenamefont {Holloway}\ \emph {et~al.}(2017)\citenamefont {Holloway}, \citenamefont {Simons}, \citenamefont {Gordon}, \citenamefont {Wilson}, \citenamefont {Cooke}, \citenamefont {Anderson},\ and\ \citenamefont {Raithel}}]{Holloway2017}%
  \BibitemOpen
  \bibfield  {author} {\bibinfo {author} {\bibfnamefont {C.~L.}\ \bibnamefont {Holloway}}, \bibinfo {author} {\bibfnamefont {M.~T.}\ \bibnamefont {Simons}}, \bibinfo {author} {\bibfnamefont {J.~A.}\ \bibnamefont {Gordon}}, \bibinfo {author} {\bibfnamefont {P.~F.}\ \bibnamefont {Wilson}}, \bibinfo {author} {\bibfnamefont {C.~M.}\ \bibnamefont {Cooke}}, \bibinfo {author} {\bibfnamefont {D.~A.}\ \bibnamefont {Anderson}},\ and\ \bibinfo {author} {\bibfnamefont {G.}~\bibnamefont {Raithel}},\ }\bibfield  {title} {\bibinfo {title} {Atom-based rf electric field metrology: From self-calibrated measurements to subwavelength and near-field imaging},\ }\href {https://doi.org/10.1109/TEMC.2016.2644616} {\bibfield  {journal} {\bibinfo  {journal} {IEEE Transactions on Electromagnetic Compatibility}\ }\textbf {\bibinfo {volume} {59}},\ \bibinfo {pages} {717} (\bibinfo {year} {2017})}\BibitemShut {NoStop}%
\bibitem [{\citenamefont {Adams}\ \emph {et~al.}(2019)\citenamefont {Adams}, \citenamefont {Pritchard},\ and\ \citenamefont {Shaffer}}]{Adams_2019}%
  \BibitemOpen
  \bibfield  {author} {\bibinfo {author} {\bibfnamefont {C.~S.}\ \bibnamefont {Adams}}, \bibinfo {author} {\bibfnamefont {J.~D.}\ \bibnamefont {Pritchard}},\ and\ \bibinfo {author} {\bibfnamefont {J.~P.}\ \bibnamefont {Shaffer}},\ }\bibfield  {title} {\bibinfo {title} {Rydberg atom quantum technologies},\ }\href {https://doi.org/10.1088/1361-6455/ab52ef} {\bibfield  {journal} {\bibinfo  {journal} {Journal of Physics B: Atomic, Molecular and Optical Physics}\ }\textbf {\bibinfo {volume} {53}},\ \bibinfo {pages} {012002} (\bibinfo {year} {2019})}\BibitemShut {NoStop}%
\bibitem [{\citenamefont {Sedlacek}\ \emph {et~al.}(2013)\citenamefont {Sedlacek}, \citenamefont {Schwettmann}, \citenamefont {K\"ubler},\ and\ \citenamefont {Shaffer}}]{Sedlacek2013}%
  \BibitemOpen
  \bibfield  {author} {\bibinfo {author} {\bibfnamefont {J.~A.}\ \bibnamefont {Sedlacek}}, \bibinfo {author} {\bibfnamefont {A.}~\bibnamefont {Schwettmann}}, \bibinfo {author} {\bibfnamefont {H.}~\bibnamefont {K\"ubler}},\ and\ \bibinfo {author} {\bibfnamefont {J.~P.}\ \bibnamefont {Shaffer}},\ }\bibfield  {title} {\bibinfo {title} {Atom-based vector microwave electrometry using rubidium rydberg atoms in a vapor cell},\ }\href {https://doi.org/10.1103/PhysRevLett.111.063001} {\bibfield  {journal} {\bibinfo  {journal} {Phys. Rev. Lett.}\ }\textbf {\bibinfo {volume} {111}},\ \bibinfo {pages} {063001} (\bibinfo {year} {2013})}\BibitemShut {NoStop}%
\bibitem [{\citenamefont {Fan}\ \emph {et~al.}(2015)\citenamefont {Fan}, \citenamefont {Kumar}, \citenamefont {Sedlacek}, \citenamefont {Kübler}, \citenamefont {Karimkashi},\ and\ \citenamefont {Shaffer}}]{Fan_2015}%
  \BibitemOpen
  \bibfield  {author} {\bibinfo {author} {\bibfnamefont {H.}~\bibnamefont {Fan}}, \bibinfo {author} {\bibfnamefont {S.}~\bibnamefont {Kumar}}, \bibinfo {author} {\bibfnamefont {J.}~\bibnamefont {Sedlacek}}, \bibinfo {author} {\bibfnamefont {H.}~\bibnamefont {Kübler}}, \bibinfo {author} {\bibfnamefont {S.}~\bibnamefont {Karimkashi}},\ and\ \bibinfo {author} {\bibfnamefont {J.~P.}\ \bibnamefont {Shaffer}},\ }\bibfield  {title} {\bibinfo {title} {Atom based {RF} electric field sensing},\ }\href {https://doi.org/10.1088/0953-4075/48/20/202001} {\bibfield  {journal} {\bibinfo  {journal} {Journal of Physics B: Atomic, Molecular and Optical Physics}\ }\textbf {\bibinfo {volume} {48}},\ \bibinfo {pages} {202001} (\bibinfo {year} {2015})}\BibitemShut {NoStop}%
\bibitem [{\citenamefont {Noaman}\ \emph {et~al.}(2023)\citenamefont {Noaman}, \citenamefont {Amarloo}, \citenamefont {Pandiyan}, \citenamefont {Bobbara}, \citenamefont {Mirzaee}, \citenamefont {Nickerson}, \citenamefont {Liu}, \citenamefont {Booth},\ and\ \citenamefont {Shaffer}}]{Noaman2023}%
  \BibitemOpen
  \bibfield  {author} {\bibinfo {author} {\bibfnamefont {M.}~\bibnamefont {Noaman}}, \bibinfo {author} {\bibfnamefont {H.}~\bibnamefont {Amarloo}}, \bibinfo {author} {\bibfnamefont {R.}~\bibnamefont {Pandiyan}}, \bibinfo {author} {\bibfnamefont {S.}~\bibnamefont {Bobbara}}, \bibinfo {author} {\bibfnamefont {S.}~\bibnamefont {Mirzaee}}, \bibinfo {author} {\bibfnamefont {K.}~\bibnamefont {Nickerson}}, \bibinfo {author} {\bibfnamefont {C.}~\bibnamefont {Liu}}, \bibinfo {author} {\bibfnamefont {D.}~\bibnamefont {Booth}},\ and\ \bibinfo {author} {\bibfnamefont {J.~P.}\ \bibnamefont {Shaffer}},\ }\bibfield  {title} {\bibinfo {title} {{Vapor cell characterization and optimization for applications in Rydberg atom-based radio frequency sensing}},\ }in\ \href {https://doi.org/10.1117/12.2657184} {\emph {\bibinfo {booktitle} {Quantum Sensing, Imaging, and Precision Metrology}}},\ Vol.\ \bibinfo {volume} {12447},\ \bibinfo {editor} {edited by\ \bibinfo {editor} {\bibfnamefont {J.}~\bibnamefont {Scheuer}}\ and\ \bibinfo
  {editor} {\bibfnamefont {S.~M.}\ \bibnamefont {Shahriar}}},\ \bibinfo {organization} {International Society for Optics and Photonics}\ (\bibinfo  {publisher} {SPIE},\ \bibinfo {year} {2023})\ p.\ \bibinfo {pages} {124470V}\BibitemShut {NoStop}%
\bibitem [{\citenamefont {Simons}\ \emph {et~al.}(2019)\citenamefont {Simons}, \citenamefont {Haddab}, \citenamefont {Gordon},\ and\ \citenamefont {Holloway}}]{Simons2019}%
  \BibitemOpen
  \bibfield  {author} {\bibinfo {author} {\bibfnamefont {M.~T.}\ \bibnamefont {Simons}}, \bibinfo {author} {\bibfnamefont {A.~H.}\ \bibnamefont {Haddab}}, \bibinfo {author} {\bibfnamefont {J.~A.}\ \bibnamefont {Gordon}},\ and\ \bibinfo {author} {\bibfnamefont {C.~L.}\ \bibnamefont {Holloway}},\ }\bibfield  {title} {\bibinfo {title} {A rydberg atom-based mixer: Measuring the phase of a radio frequency wave},\ }\href {https://doi.org/10.1063/1.5088821} {\bibfield  {journal} {\bibinfo  {journal} {Applied Physics Letters}\ }\textbf {\bibinfo {volume} {114}},\ \bibinfo {pages} {114101} (\bibinfo {year} {2019})}\BibitemShut {NoStop}%
\bibitem [{\citenamefont {Robinson}\ \emph {et~al.}(2021)\citenamefont {Robinson}, \citenamefont {Prajapati}, \citenamefont {Senic}, \citenamefont {Simons},\ and\ \citenamefont {Holloway}}]{Holloway21AngleOfArrival}%
  \BibitemOpen
  \bibfield  {author} {\bibinfo {author} {\bibfnamefont {A.~K.}\ \bibnamefont {Robinson}}, \bibinfo {author} {\bibfnamefont {N.}~\bibnamefont {Prajapati}}, \bibinfo {author} {\bibfnamefont {D.}~\bibnamefont {Senic}}, \bibinfo {author} {\bibfnamefont {M.~T.}\ \bibnamefont {Simons}},\ and\ \bibinfo {author} {\bibfnamefont {C.~L.}\ \bibnamefont {Holloway}},\ }\bibfield  {title} {\bibinfo {title} {Determining the angle-of-arrival of a radio-frequency source with a rydberg atom-based sensor},\ }\href {https://doi.org/10.1063/5.0045601} {\bibfield  {journal} {\bibinfo  {journal} {Applied Physics Letters}\ }\textbf {\bibinfo {volume} {118}},\ \bibinfo {pages} {114001} (\bibinfo {year} {2021})}\BibitemShut {NoStop}%
\bibitem [{\citenamefont {Schlossberger}\ \emph {et~al.}(2025)\citenamefont {Schlossberger}, \citenamefont {Talashila}, \citenamefont {Prajapati},\ and\ \citenamefont {Holloway}}]{schlossberger2025angleofarrival}%
  \BibitemOpen
  \bibfield  {author} {\bibinfo {author} {\bibfnamefont {N.}~\bibnamefont {Schlossberger}}, \bibinfo {author} {\bibfnamefont {R.}~\bibnamefont {Talashila}}, \bibinfo {author} {\bibfnamefont {N.}~\bibnamefont {Prajapati}},\ and\ \bibinfo {author} {\bibfnamefont {C.~L.}\ \bibnamefont {Holloway}},\ }\href@noop {} {\bibinfo {title} {Angle-of-arrival detection of radio-frequency waves via rydberg atom fluorescence imaging of standing waves in a glass vapor cell}} (\bibinfo {year} {2025}),\ \Eprint {https://arxiv.org/abs/2504.18028} {arXiv:2504.18028 [physics.atom-ph]} \BibitemShut {NoStop}%
\bibitem [{\citenamefont {Meyer}\ \emph {et~al.}(2018)\citenamefont {Meyer}, \citenamefont {Cox}, \citenamefont {Fatemi},\ and\ \citenamefont {Kunz}}]{Meyer2018}%
  \BibitemOpen
  \bibfield  {author} {\bibinfo {author} {\bibfnamefont {D.~H.}\ \bibnamefont {Meyer}}, \bibinfo {author} {\bibfnamefont {K.~C.}\ \bibnamefont {Cox}}, \bibinfo {author} {\bibfnamefont {F.~K.}\ \bibnamefont {Fatemi}},\ and\ \bibinfo {author} {\bibfnamefont {P.~D.}\ \bibnamefont {Kunz}},\ }\bibfield  {title} {\bibinfo {title} {Digital communication with rydberg atoms and amplitude-modulated microwave fields},\ }\href {https://doi.org/10.1063/1.5028357} {\bibfield  {journal} {\bibinfo  {journal} {Applied Physics Letters}\ }\textbf {\bibinfo {volume} {112}},\ \bibinfo {pages} {211108} (\bibinfo {year} {2018})}\BibitemShut {NoStop}%
\bibitem [{\citenamefont {Anderson}\ \emph {et~al.}(2021)\citenamefont {Anderson}, \citenamefont {Sapiro},\ and\ \citenamefont {Raithel}}]{Anderson2021}%
  \BibitemOpen
  \bibfield  {author} {\bibinfo {author} {\bibfnamefont {D.~A.}\ \bibnamefont {Anderson}}, \bibinfo {author} {\bibfnamefont {R.~E.}\ \bibnamefont {Sapiro}},\ and\ \bibinfo {author} {\bibfnamefont {G.}~\bibnamefont {Raithel}},\ }\bibfield  {title} {\bibinfo {title} {An atomic receiver for am and fm radio communication},\ }\href {https://doi.org/10.1109/TAP.2020.2987112} {\bibfield  {journal} {\bibinfo  {journal} {IEEE Transactions on Antennas and Propagation}\ }\textbf {\bibinfo {volume} {69}},\ \bibinfo {pages} {2455} (\bibinfo {year} {2021})}\BibitemShut {NoStop}%
\bibitem [{\citenamefont {Jiao}\ \emph {et~al.}(2019)\citenamefont {Jiao}, \citenamefont {Han}, \citenamefont {Fan}, \citenamefont {Raithel}, \citenamefont {Zhao},\ and\ \citenamefont {Jia}}]{Jiao_2019}%
  \BibitemOpen
  \bibfield  {author} {\bibinfo {author} {\bibfnamefont {Y.}~\bibnamefont {Jiao}}, \bibinfo {author} {\bibfnamefont {X.}~\bibnamefont {Han}}, \bibinfo {author} {\bibfnamefont {J.}~\bibnamefont {Fan}}, \bibinfo {author} {\bibfnamefont {G.}~\bibnamefont {Raithel}}, \bibinfo {author} {\bibfnamefont {J.}~\bibnamefont {Zhao}},\ and\ \bibinfo {author} {\bibfnamefont {S.}~\bibnamefont {Jia}},\ }\bibfield  {title} {\bibinfo {title} {Atom-based receiver for amplitude-modulated baseband signals in high-frequency radio communication},\ }\href {https://doi.org/10.7567/1882-0786/ab5463} {\bibfield  {journal} {\bibinfo  {journal} {Applied Physics Express}\ }\textbf {\bibinfo {volume} {12}},\ \bibinfo {pages} {126002} (\bibinfo {year} {2019})}\BibitemShut {NoStop}%
\bibitem [{\citenamefont {Holloway}\ \emph {et~al.}(2021)\citenamefont {Holloway}, \citenamefont {Simons}, \citenamefont {Haddab}, \citenamefont {Gordon}, \citenamefont {Anderson}, \citenamefont {Raithel},\ and\ \citenamefont {Voran}}]{Holloway2021AMR}%
  \BibitemOpen
  \bibfield  {author} {\bibinfo {author} {\bibfnamefont {C.}~\bibnamefont {Holloway}}, \bibinfo {author} {\bibfnamefont {M.}~\bibnamefont {Simons}}, \bibinfo {author} {\bibfnamefont {A.~H.}\ \bibnamefont {Haddab}}, \bibinfo {author} {\bibfnamefont {J.~A.}\ \bibnamefont {Gordon}}, \bibinfo {author} {\bibfnamefont {D.~A.}\ \bibnamefont {Anderson}}, \bibinfo {author} {\bibfnamefont {G.}~\bibnamefont {Raithel}},\ and\ \bibinfo {author} {\bibfnamefont {S.}~\bibnamefont {Voran}},\ }\bibfield  {title} {\bibinfo {title} {A multiple-band rydberg atom-based receiver: Am/fm stereo reception},\ }\href {https://doi.org/10.1109/MAP.2020.2976914} {\bibfield  {journal} {\bibinfo  {journal} {IEEE Antennas and Propagation Magazine}\ }\textbf {\bibinfo {volume} {63}},\ \bibinfo {pages} {63} (\bibinfo {year} {2021})}\BibitemShut {NoStop}%
\bibitem [{\citenamefont {Song}\ \emph {et~al.}(2019)\citenamefont {Song}, \citenamefont {Liu}, \citenamefont {Liu}, \citenamefont {Zhang}, \citenamefont {Zou}, \citenamefont {Zhang},\ and\ \citenamefont {Qu}}]{Song2019}%
  \BibitemOpen
  \bibfield  {author} {\bibinfo {author} {\bibfnamefont {Z.}~\bibnamefont {Song}}, \bibinfo {author} {\bibfnamefont {H.}~\bibnamefont {Liu}}, \bibinfo {author} {\bibfnamefont {X.}~\bibnamefont {Liu}}, \bibinfo {author} {\bibfnamefont {W.}~\bibnamefont {Zhang}}, \bibinfo {author} {\bibfnamefont {H.}~\bibnamefont {Zou}}, \bibinfo {author} {\bibfnamefont {J.}~\bibnamefont {Zhang}},\ and\ \bibinfo {author} {\bibfnamefont {J.}~\bibnamefont {Qu}},\ }\bibfield  {title} {\bibinfo {title} {Rydberg-atom-based digital communication using a continuously tunable radio-frequency carrier},\ }\href {https://doi.org/10.1364/OE.27.008848} {\bibfield  {journal} {\bibinfo  {journal} {Opt. Express}\ }\textbf {\bibinfo {volume} {27}},\ \bibinfo {pages} {8848} (\bibinfo {year} {2019})}\BibitemShut {NoStop}%
\bibitem [{\citenamefont {Bohaichuk}\ \emph {et~al.}(2022)\citenamefont {Bohaichuk}, \citenamefont {Booth}, \citenamefont {Nickerson}, \citenamefont {Tai},\ and\ \citenamefont {Shaffer}}]{Bohaichuk2022}%
  \BibitemOpen
  \bibfield  {author} {\bibinfo {author} {\bibfnamefont {S.~M.}\ \bibnamefont {Bohaichuk}}, \bibinfo {author} {\bibfnamefont {D.}~\bibnamefont {Booth}}, \bibinfo {author} {\bibfnamefont {K.}~\bibnamefont {Nickerson}}, \bibinfo {author} {\bibfnamefont {H.}~\bibnamefont {Tai}},\ and\ \bibinfo {author} {\bibfnamefont {J.~P.}\ \bibnamefont {Shaffer}},\ }\bibfield  {title} {\bibinfo {title} {Origins of rydberg-atom electrometer transient response and its impact on radio-frequency pulse sensing},\ }\href {https://doi.org/10.1103/PhysRevApplied.18.034030} {\bibfield  {journal} {\bibinfo  {journal} {Phys. Rev. Appl.}\ }\textbf {\bibinfo {volume} {18}},\ \bibinfo {pages} {034030} (\bibinfo {year} {2022})}\BibitemShut {NoStop}%
\bibitem [{\citenamefont {Morigi}\ \emph {et~al.}(2002)\citenamefont {Morigi}, \citenamefont {Franke-Arnold},\ and\ \citenamefont {Oppo}}]{Morigi02}%
  \BibitemOpen
  \bibfield  {author} {\bibinfo {author} {\bibfnamefont {G.}~\bibnamefont {Morigi}}, \bibinfo {author} {\bibfnamefont {S.}~\bibnamefont {Franke-Arnold}},\ and\ \bibinfo {author} {\bibfnamefont {G.-L.}\ \bibnamefont {Oppo}},\ }\bibfield  {title} {\bibinfo {title} {Phase-dependent interaction in a four-level atomic configuration},\ }\href {https://doi.org/10.1103/PhysRevA.66.053409} {\bibfield  {journal} {\bibinfo  {journal} {Phys. Rev. A}\ }\textbf {\bibinfo {volume} {66}},\ \bibinfo {pages} {053409} (\bibinfo {year} {2002})}\BibitemShut {NoStop}%
\bibitem [{\citenamefont {Kajari-Schr\"oder}\ \emph {et~al.}(2007)\citenamefont {Kajari-Schr\"oder}, \citenamefont {Morigi}, \citenamefont {Franke-Arnold},\ and\ \citenamefont {Oppo}}]{Morigi07PhaseVapor}%
  \BibitemOpen
  \bibfield  {author} {\bibinfo {author} {\bibfnamefont {S.}~\bibnamefont {Kajari-Schr\"oder}}, \bibinfo {author} {\bibfnamefont {G.}~\bibnamefont {Morigi}}, \bibinfo {author} {\bibfnamefont {S.}~\bibnamefont {Franke-Arnold}},\ and\ \bibinfo {author} {\bibfnamefont {G.-L.}\ \bibnamefont {Oppo}},\ }\bibfield  {title} {\bibinfo {title} {Phase-dependent light propagation in atomic vapors},\ }\href {https://doi.org/10.1103/PhysRevA.75.013816} {\bibfield  {journal} {\bibinfo  {journal} {Phys. Rev. A}\ }\textbf {\bibinfo {volume} {75}},\ \bibinfo {pages} {013816} (\bibinfo {year} {2007})}\BibitemShut {NoStop}%
\bibitem [{\citenamefont {Anderson}\ \emph {et~al.}(2022)\citenamefont {Anderson}, \citenamefont {Sapiro}, \citenamefont {Gon\ifmmode~\mbox{\c{c}}\else \c{c}\fi{}alves}, \citenamefont {Cardman},\ and\ \citenamefont {Raithel}}]{RaithelPhase22}%
  \BibitemOpen
  \bibfield  {author} {\bibinfo {author} {\bibfnamefont {D.}~\bibnamefont {Anderson}}, \bibinfo {author} {\bibfnamefont {R.}~\bibnamefont {Sapiro}}, \bibinfo {author} {\bibfnamefont {L.}~\bibnamefont {Gon\ifmmode~\mbox{\c{c}}\else \c{c}\fi{}alves}}, \bibinfo {author} {\bibfnamefont {R.}~\bibnamefont {Cardman}},\ and\ \bibinfo {author} {\bibfnamefont {G.}~\bibnamefont {Raithel}},\ }\bibfield  {title} {\bibinfo {title} {Optical radio-frequency phase measurement with an internal-state rydberg atom interferometer},\ }\href {https://doi.org/10.1103/PhysRevApplied.17.044020} {\bibfield  {journal} {\bibinfo  {journal} {Phys. Rev. Appl.}\ }\textbf {\bibinfo {volume} {17}},\ \bibinfo {pages} {044020} (\bibinfo {year} {2022})}\BibitemShut {NoStop}%
\bibitem [{\citenamefont {Berweger}\ \emph {et~al.}(2023)\citenamefont {Berweger}, \citenamefont {Artusio-Glimpse}, \citenamefont {Rotunno}, \citenamefont {Prajapati}, \citenamefont {Christesen}, \citenamefont {Moore}, \citenamefont {Simons},\ and\ \citenamefont {Holloway}}]{HollowayPhase23}%
  \BibitemOpen
  \bibfield  {author} {\bibinfo {author} {\bibfnamefont {S.}~\bibnamefont {Berweger}}, \bibinfo {author} {\bibfnamefont {A.~B.}\ \bibnamefont {Artusio-Glimpse}}, \bibinfo {author} {\bibfnamefont {A.~P.}\ \bibnamefont {Rotunno}}, \bibinfo {author} {\bibfnamefont {N.}~\bibnamefont {Prajapati}}, \bibinfo {author} {\bibfnamefont {J.~D.}\ \bibnamefont {Christesen}}, \bibinfo {author} {\bibfnamefont {K.~R.}\ \bibnamefont {Moore}}, \bibinfo {author} {\bibfnamefont {M.~T.}\ \bibnamefont {Simons}},\ and\ \bibinfo {author} {\bibfnamefont {C.~L.}\ \bibnamefont {Holloway}},\ }\bibfield  {title} {\bibinfo {title} {Closed-loop quantum interferometry for phase-resolved rydberg-atom field sensing},\ }\href {https://doi.org/10.1103/PhysRevApplied.20.054009} {\bibfield  {journal} {\bibinfo  {journal} {Phys. Rev. Appl.}\ }\textbf {\bibinfo {volume} {20}},\ \bibinfo {pages} {054009} (\bibinfo {year} {2023})}\BibitemShut {NoStop}%
\bibitem [{\citenamefont {Borówka}\ \emph {et~al.}(2025)\citenamefont {Borówka}, \citenamefont {Mazelanik}, \citenamefont {Wasilewski},\ and\ \citenamefont {Parniak}}]{borowka2025PhaseEp}%
  \BibitemOpen
  \bibfield  {author} {\bibinfo {author} {\bibfnamefont {S.}~\bibnamefont {Borówka}}, \bibinfo {author} {\bibfnamefont {M.}~\bibnamefont {Mazelanik}}, \bibinfo {author} {\bibfnamefont {W.}~\bibnamefont {Wasilewski}},\ and\ \bibinfo {author} {\bibfnamefont {M.}~\bibnamefont {Parniak}},\ }\href@noop {} {\bibinfo {title} {Optically-biased rydberg microwave receiver enabled by hybrid nonlinear interferometry}} (\bibinfo {year} {2025}),\ \Eprint {https://arxiv.org/abs/2403.05310} {arXiv:2403.05310 [physics.atom-ph]} \BibitemShut {NoStop}%
\bibitem [{\citenamefont {Kasza}\ \emph {et~al.}(2024)\citenamefont {Kasza}, \citenamefont {Borówka}, \citenamefont {Wasilewski},\ and\ \citenamefont {Parniak}}]{kasza2024PhaseTheory}%
  \BibitemOpen
  \bibfield  {author} {\bibinfo {author} {\bibfnamefont {B.}~\bibnamefont {Kasza}}, \bibinfo {author} {\bibfnamefont {S.}~\bibnamefont {Borówka}}, \bibinfo {author} {\bibfnamefont {W.}~\bibnamefont {Wasilewski}},\ and\ \bibinfo {author} {\bibfnamefont {M.}~\bibnamefont {Parniak}},\ }\href@noop {} {\bibinfo {title} {Atomic-optical interferometry in fractured loops: a general solution for rydberg radio frequency receivers}} (\bibinfo {year} {2024}),\ \Eprint {https://arxiv.org/abs/2412.07632} {arXiv:2412.07632 [physics.atom-ph]} \BibitemShut {NoStop}%
\bibitem [{\citenamefont {Bohaichuk}\ \emph {et~al.}(2023)\citenamefont {Bohaichuk}, \citenamefont {Ripka}, \citenamefont {Venu}, \citenamefont {Christaller}, \citenamefont {Liu}, \citenamefont {Schmidt}, \citenamefont {K\"ubler},\ and\ \citenamefont {Shaffer}}]{Bohaichuk2023threephoton}%
  \BibitemOpen
  \bibfield  {author} {\bibinfo {author} {\bibfnamefont {S.~M.}\ \bibnamefont {Bohaichuk}}, \bibinfo {author} {\bibfnamefont {F.}~\bibnamefont {Ripka}}, \bibinfo {author} {\bibfnamefont {V.}~\bibnamefont {Venu}}, \bibinfo {author} {\bibfnamefont {F.}~\bibnamefont {Christaller}}, \bibinfo {author} {\bibfnamefont {C.}~\bibnamefont {Liu}}, \bibinfo {author} {\bibfnamefont {M.}~\bibnamefont {Schmidt}}, \bibinfo {author} {\bibfnamefont {H.}~\bibnamefont {K\"ubler}},\ and\ \bibinfo {author} {\bibfnamefont {J.~P.}\ \bibnamefont {Shaffer}},\ }\bibfield  {title} {\bibinfo {title} {Three-photon rydberg-atom-based radio-frequency sensing scheme with narrow linewidth},\ }\href {https://doi.org/10.1103/PhysRevApplied.20.L061004} {\bibfield  {journal} {\bibinfo  {journal} {Phys. Rev. Appl.}\ }\textbf {\bibinfo {volume} {20}},\ \bibinfo {pages} {L061004} (\bibinfo {year} {2023})}\BibitemShut {NoStop}%
\bibitem [{\citenamefont {Venu}\ \emph {et~al.}(2025)\citenamefont {Venu}, \citenamefont {Bohaichuk}, \citenamefont {Christaller}, \citenamefont {Schmidt}, \citenamefont {K{\"u}bler},\ and\ \citenamefont {Shaffer}}]{Venu2025}%
  \BibitemOpen
  \bibfield  {author} {\bibinfo {author} {\bibfnamefont {V.}~\bibnamefont {Venu}}, \bibinfo {author} {\bibfnamefont {S.~M.}\ \bibnamefont {Bohaichuk}}, \bibinfo {author} {\bibfnamefont {F.}~\bibnamefont {Christaller}}, \bibinfo {author} {\bibfnamefont {M.}~\bibnamefont {Schmidt}}, \bibinfo {author} {\bibfnamefont {H.}~\bibnamefont {K{\"u}bler}},\ and\ \bibinfo {author} {\bibfnamefont {J.~P.}\ \bibnamefont {Shaffer}},\ }\bibfield  {title} {\bibinfo {title} {{Three-photon Rydberg atom electrometry with enhanced sensitivity}},\ }in\ \href {https://doi.org/10.1117/12.3053172} {\emph {\bibinfo {booktitle} {Quantum Sensing, Imaging, and Precision Metrology III}}},\ Vol.\ \bibinfo {volume} {13392},\ \bibinfo {editor} {edited by\ \bibinfo {editor} {\bibfnamefont {S.~M.}\ \bibnamefont {Shahriar}}},\ \bibinfo {organization} {International Society for Optics and Photonics}\ (\bibinfo  {publisher} {SPIE},\ \bibinfo {year} {2025})\ p.\ \bibinfo {pages} {1339216}\BibitemShut {NoStop}%
\bibitem [{\citenamefont {Gordon}\ \emph {et~al.}(2014)\citenamefont {Gordon}, \citenamefont {Holloway}, \citenamefont {Schwarzkopf}, \citenamefont {Anderson}, \citenamefont {Miller}, \citenamefont {Thaicharoen},\ and\ \citenamefont {Raithel}}]{Gordon2014}%
  \BibitemOpen
  \bibfield  {author} {\bibinfo {author} {\bibfnamefont {J.~A.}\ \bibnamefont {Gordon}}, \bibinfo {author} {\bibfnamefont {C.~L.}\ \bibnamefont {Holloway}}, \bibinfo {author} {\bibfnamefont {A.}~\bibnamefont {Schwarzkopf}}, \bibinfo {author} {\bibfnamefont {D.~A.}\ \bibnamefont {Anderson}}, \bibinfo {author} {\bibfnamefont {S.}~\bibnamefont {Miller}}, \bibinfo {author} {\bibfnamefont {N.}~\bibnamefont {Thaicharoen}},\ and\ \bibinfo {author} {\bibfnamefont {G.}~\bibnamefont {Raithel}},\ }\bibfield  {title} {\bibinfo {title} {Millimeter wave detection via autler-townes splitting in rubidium rydberg atoms},\ }\href {https://doi.org/10.1063/1.4890094} {\bibfield  {journal} {\bibinfo  {journal} {Applied Physics Letters}\ }\textbf {\bibinfo {volume} {105}},\ \bibinfo {pages} {024104} (\bibinfo {year} {2014})}\BibitemShut {NoStop}%
\bibitem [{\citenamefont {Schmidt}\ \emph {et~al.}(2024)\citenamefont {Schmidt}, \citenamefont {Bohaichuk}, \citenamefont {Venu}, \citenamefont {Christaller}, \citenamefont {Liu}, \citenamefont {Ripka}, \citenamefont {K\"{u}bler},\ and\ \citenamefont {Shaffer}}]{Schmidt2024}%
  \BibitemOpen
  \bibfield  {author} {\bibinfo {author} {\bibfnamefont {M.}~\bibnamefont {Schmidt}}, \bibinfo {author} {\bibfnamefont {S.}~\bibnamefont {Bohaichuk}}, \bibinfo {author} {\bibfnamefont {V.}~\bibnamefont {Venu}}, \bibinfo {author} {\bibfnamefont {F.}~\bibnamefont {Christaller}}, \bibinfo {author} {\bibfnamefont {C.}~\bibnamefont {Liu}}, \bibinfo {author} {\bibfnamefont {F.}~\bibnamefont {Ripka}}, \bibinfo {author} {\bibfnamefont {H.}~\bibnamefont {K\"{u}bler}},\ and\ \bibinfo {author} {\bibfnamefont {J.~P.}\ \bibnamefont {Shaffer}},\ }\bibfield  {title} {\bibinfo {title} {Rydberg-atom-based radio-frequency sensors: amplitude-regime sensing},\ }\href {https://doi.org/10.1364/OE.530148} {\bibfield  {journal} {\bibinfo  {journal} {Opt. Express}\ }\textbf {\bibinfo {volume} {32}},\ \bibinfo {pages} {27768} (\bibinfo {year} {2024})}\BibitemShut {NoStop}%
\bibitem [{\citenamefont {Shaffer}\ and\ \citenamefont {Kübler}(2018)}]{Kuebler2018}%
  \BibitemOpen
  \bibfield  {author} {\bibinfo {author} {\bibfnamefont {J.~P.}\ \bibnamefont {Shaffer}}\ and\ \bibinfo {author} {\bibfnamefont {H.}~\bibnamefont {Kübler}},\ }\bibfield  {title} {\bibinfo {title} {{A read-out enhancement for microwave electric field sensing with Rydberg atoms}},\ }\href {https://doi.org/10.1117/12.2309386} {\bibfield  {journal} {\bibinfo  {journal} {SPIE}\ }\textbf {\bibinfo {volume} {10674}},\ \bibinfo {pages} {39 } (\bibinfo {year} {2018})}\BibitemShut {NoStop}%
\bibitem [{\citenamefont {Kumar}\ \emph {et~al.}(2016)\citenamefont {Kumar}, \citenamefont {Fan}, \citenamefont {Kübler}, \citenamefont {Sheng},\ and\ \citenamefont {Shaffer}}]{Kumar2016}%
  \BibitemOpen
  \bibfield  {author} {\bibinfo {author} {\bibfnamefont {S.}~\bibnamefont {Kumar}}, \bibinfo {author} {\bibfnamefont {H.}~\bibnamefont {Fan}}, \bibinfo {author} {\bibfnamefont {H.}~\bibnamefont {Kübler}}, \bibinfo {author} {\bibfnamefont {J.}~\bibnamefont {Sheng}},\ and\ \bibinfo {author} {\bibfnamefont {J.~P.}\ \bibnamefont {Shaffer}},\ }\bibfield  {title} {\bibinfo {title} {Atom-based sensing of weak radio frequency electric fields using homodyne readout},\ }\href {https://doi.org/10.1038/srep42981} {\bibfield  {journal} {\bibinfo  {journal} {Scientific Reports}\ }\textbf {\bibinfo {volume} {7}} (\bibinfo {year} {2016})}\BibitemShut {NoStop}%
\bibitem [{\citenamefont {Kumar}\ \emph {et~al.}(2017)\citenamefont {Kumar}, \citenamefont {Fan}, \citenamefont {K\"{u}bler}, \citenamefont {Jahangiri},\ and\ \citenamefont {Shaffer}}]{Kumar2017}%
  \BibitemOpen
  \bibfield  {author} {\bibinfo {author} {\bibfnamefont {S.}~\bibnamefont {Kumar}}, \bibinfo {author} {\bibfnamefont {H.}~\bibnamefont {Fan}}, \bibinfo {author} {\bibfnamefont {H.}~\bibnamefont {K\"{u}bler}}, \bibinfo {author} {\bibfnamefont {A.~J.}\ \bibnamefont {Jahangiri}},\ and\ \bibinfo {author} {\bibfnamefont {J.~P.}\ \bibnamefont {Shaffer}},\ }\bibfield  {title} {\bibinfo {title} {Rydberg-atom based radio-frequency electrometry using frequency modulation spectroscopy in room temperature vapor cells},\ }\href {https://doi.org/10.1364/OE.25.008625} {\bibfield  {journal} {\bibinfo  {journal} {Opt. Express}\ }\textbf {\bibinfo {volume} {25}},\ \bibinfo {pages} {8625} (\bibinfo {year} {2017})}\BibitemShut {NoStop}%
\bibitem [{\citenamefont {Sapiro}\ \emph {et~al.}(2020)\citenamefont {Sapiro}, \citenamefont {Raithel},\ and\ \citenamefont {Anderson}}]{Sapiro_2020}%
  \BibitemOpen
  \bibfield  {author} {\bibinfo {author} {\bibfnamefont {R.~E.}\ \bibnamefont {Sapiro}}, \bibinfo {author} {\bibfnamefont {G.}~\bibnamefont {Raithel}},\ and\ \bibinfo {author} {\bibfnamefont {D.~A.}\ \bibnamefont {Anderson}},\ }\bibfield  {title} {\bibinfo {title} {Time dependence of rydberg {EIT} in pulsed optical and {RF} fields},\ }\href {https://doi.org/10.1088/1361-6455/ab7426} {\bibfield  {journal} {\bibinfo  {journal} {Journal of Physics B: Atomic, Molecular and Optical Physics}\ }\textbf {\bibinfo {volume} {53}},\ \bibinfo {pages} {094003} (\bibinfo {year} {2020})}\BibitemShut {NoStop}%
\bibitem [{\citenamefont {Cai}\ \emph {et~al.}(2022)\citenamefont {Cai}, \citenamefont {Xu}, \citenamefont {You},\ and\ \citenamefont {Liu}}]{Cai2022}%
  \BibitemOpen
  \bibfield  {author} {\bibinfo {author} {\bibfnamefont {M.}~\bibnamefont {Cai}}, \bibinfo {author} {\bibfnamefont {Z.}~\bibnamefont {Xu}}, \bibinfo {author} {\bibfnamefont {S.}~\bibnamefont {You}},\ and\ \bibinfo {author} {\bibfnamefont {H.}~\bibnamefont {Liu}},\ }\bibfield  {title} {\bibinfo {title} {Sensitivity improvement and determination of rydberg atom-based microwave sensor},\ }\bibfield  {journal} {\bibinfo  {journal} {Photonics}\ }\textbf {\bibinfo {volume} {9}},\ \href {https://doi.org/10.3390/photonics9040250} {10.3390/photonics9040250} (\bibinfo {year} {2022})\BibitemShut {NoStop}%
\bibitem [{\citenamefont {Liu}\ \emph {et~al.}(2022)\citenamefont {Liu}, \citenamefont {Zhang}, \citenamefont {Liu}, \citenamefont {Zhang}, \citenamefont {Zhu}, \citenamefont {Gao}, \citenamefont {Guo}, \citenamefont {Ding},\ and\ \citenamefont {Shi}}]{Bang2022}%
  \BibitemOpen
  \bibfield  {author} {\bibinfo {author} {\bibfnamefont {B.}~\bibnamefont {Liu}}, \bibinfo {author} {\bibfnamefont {L.-H.}\ \bibnamefont {Zhang}}, \bibinfo {author} {\bibfnamefont {Z.-K.}\ \bibnamefont {Liu}}, \bibinfo {author} {\bibfnamefont {Z.-Y.}\ \bibnamefont {Zhang}}, \bibinfo {author} {\bibfnamefont {Z.-H.}\ \bibnamefont {Zhu}}, \bibinfo {author} {\bibfnamefont {W.}~\bibnamefont {Gao}}, \bibinfo {author} {\bibfnamefont {G.-C.}\ \bibnamefont {Guo}}, \bibinfo {author} {\bibfnamefont {D.-S.}\ \bibnamefont {Ding}},\ and\ \bibinfo {author} {\bibfnamefont {B.-S.}\ \bibnamefont {Shi}},\ }\bibfield  {title} {\bibinfo {title} {Highly sensitive measurement of a megahertz rf electric field with a rydberg-atom sensor},\ }\href {https://doi.org/10.1103/PhysRevApplied.18.014045} {\bibfield  {journal} {\bibinfo  {journal} {Phys. Rev. Applied}\ }\textbf {\bibinfo {volume} {18}},\ \bibinfo {pages} {014045} (\bibinfo {year} {2022})}\BibitemShut {NoStop}%
\bibitem [{\citenamefont {Jing}\ \emph {et~al.}(2020)\citenamefont {Jing}, \citenamefont {Hu}, \citenamefont {Ma}, \citenamefont {Zhang}, \citenamefont {Zhang}, \citenamefont {Xiao},\ and\ \citenamefont {Jia}}]{Jing2020}%
  \BibitemOpen
  \bibfield  {author} {\bibinfo {author} {\bibfnamefont {M.}~\bibnamefont {Jing}}, \bibinfo {author} {\bibfnamefont {Y.}~\bibnamefont {Hu}}, \bibinfo {author} {\bibfnamefont {J.}~\bibnamefont {Ma}}, \bibinfo {author} {\bibfnamefont {H.}~\bibnamefont {Zhang}}, \bibinfo {author} {\bibfnamefont {L.}~\bibnamefont {Zhang}}, \bibinfo {author} {\bibfnamefont {L.}~\bibnamefont {Xiao}},\ and\ \bibinfo {author} {\bibfnamefont {S.}~\bibnamefont {Jia}},\ }\bibfield  {title} {\bibinfo {title} {Atomic superheterodyne receiver based on microwave-dressed rydberg spectroscopy},\ }\href {https://doi.org/10.1038/s41567-020-0918-5} {\bibfield  {journal} {\bibinfo  {journal} {Nature Physics}\ }\textbf {\bibinfo {volume} {16}},\ \bibinfo {pages} {911} (\bibinfo {year} {2020})}\BibitemShut {NoStop}%
\bibitem [{\citenamefont {Simons}\ \emph {et~al.}(2021)\citenamefont {Simons}, \citenamefont {Artusio-Glimpse}, \citenamefont {Robinson}, \citenamefont {Prajapati},\ and\ \citenamefont {Holloway}}]{Simons2021}%
  \BibitemOpen
  \bibfield  {author} {\bibinfo {author} {\bibfnamefont {M.~T.}\ \bibnamefont {Simons}}, \bibinfo {author} {\bibfnamefont {A.~B.}\ \bibnamefont {Artusio-Glimpse}}, \bibinfo {author} {\bibfnamefont {A.~K.}\ \bibnamefont {Robinson}}, \bibinfo {author} {\bibfnamefont {N.}~\bibnamefont {Prajapati}},\ and\ \bibinfo {author} {\bibfnamefont {C.~L.}\ \bibnamefont {Holloway}},\ }\bibfield  {title} {\bibinfo {title} {Rydberg atom-based sensors for radio-frequency electric field metrology, sensing, and communications},\ }\href {https://doi.org/https://doi.org/10.1016/j.measen.2021.100273} {\bibfield  {journal} {\bibinfo  {journal} {Measurement: Sensors}\ }\textbf {\bibinfo {volume} {18}},\ \bibinfo {pages} {100273} (\bibinfo {year} {2021})}\BibitemShut {NoStop}%
\bibitem [{\citenamefont {Cui}\ \emph {et~al.}(2023)\citenamefont {Cui}, \citenamefont {Jia}, \citenamefont {Hao}, \citenamefont {Wang}, \citenamefont {Zhou}, \citenamefont {Liu}, \citenamefont {Yu}, \citenamefont {Mei}, \citenamefont {Bai}, \citenamefont {Bao}, \citenamefont {Hu}, \citenamefont {Wang}, \citenamefont {Liu}, \citenamefont {Zhang}, \citenamefont {Xie},\ and\ \citenamefont {Zhong}}]{Cui2023}%
  \BibitemOpen
  \bibfield  {author} {\bibinfo {author} {\bibfnamefont {Y.}~\bibnamefont {Cui}}, \bibinfo {author} {\bibfnamefont {F.-D.}\ \bibnamefont {Jia}}, \bibinfo {author} {\bibfnamefont {J.-H.}\ \bibnamefont {Hao}}, \bibinfo {author} {\bibfnamefont {Y.-H.}\ \bibnamefont {Wang}}, \bibinfo {author} {\bibfnamefont {F.}~\bibnamefont {Zhou}}, \bibinfo {author} {\bibfnamefont {X.-B.}\ \bibnamefont {Liu}}, \bibinfo {author} {\bibfnamefont {Y.-H.}\ \bibnamefont {Yu}}, \bibinfo {author} {\bibfnamefont {J.}~\bibnamefont {Mei}}, \bibinfo {author} {\bibfnamefont {J.-H.}\ \bibnamefont {Bai}}, \bibinfo {author} {\bibfnamefont {Y.-Y.}\ \bibnamefont {Bao}}, \bibinfo {author} {\bibfnamefont {D.}~\bibnamefont {Hu}}, \bibinfo {author} {\bibfnamefont {Y.}~\bibnamefont {Wang}}, \bibinfo {author} {\bibfnamefont {Y.}~\bibnamefont {Liu}}, \bibinfo {author} {\bibfnamefont {J.}~\bibnamefont {Zhang}}, \bibinfo {author} {\bibfnamefont {F.}~\bibnamefont {Xie}},\ and\ \bibinfo {author} {\bibfnamefont {Z.-P.}\ \bibnamefont {Zhong}},\ }\bibfield
  {title} {\bibinfo {title} {Extending bandwidth sensitivity of rydberg-atom-based microwave electrometry using an auxiliary microwave field},\ }\href {https://doi.org/10.1103/PhysRevA.107.043102} {\bibfield  {journal} {\bibinfo  {journal} {Phys. Rev. A}\ }\textbf {\bibinfo {volume} {107}},\ \bibinfo {pages} {043102} (\bibinfo {year} {2023})}\BibitemShut {NoStop}%
\bibitem [{\citenamefont {Anderson}\ \emph {et~al.}(2020)\citenamefont {Anderson}, \citenamefont {Sapiro},\ and\ \citenamefont {Raithel}}]{Anderson2019RydbergAF}%
  \BibitemOpen
  \bibfield  {author} {\bibinfo {author} {\bibfnamefont {D.~A.}\ \bibnamefont {Anderson}}, \bibinfo {author} {\bibfnamefont {R.~E.}\ \bibnamefont {Sapiro}},\ and\ \bibinfo {author} {\bibfnamefont {G.}~\bibnamefont {Raithel}},\ }\bibfield  {title} {\bibinfo {title} {Rydberg atoms for radio-frequency communications and sensing: Atomic receivers for pulsed rf field and phase detection},\ }\href {https://doi.org/10.1109/MAES.2019.2960922} {\bibfield  {journal} {\bibinfo  {journal} {IEEE Aerospace and Electronic Systems Magazine}\ }\textbf {\bibinfo {volume} {35}},\ \bibinfo {pages} {48} (\bibinfo {year} {2020})}\BibitemShut {NoStop}%
\bibitem [{sup()}]{supp}%
  \BibitemOpen
  \href@noop {} {}\bibinfo {note} {See Supplemental Material at [URL will be inserted by publisher] for more details on the theoretical description, the weak probe approximation and bandwidth analysis.}\BibitemShut {Stop}%
\bibitem [{\citenamefont {Fleischhauer}\ \emph {et~al.}(2005)\citenamefont {Fleischhauer}, \citenamefont {Imamoglu},\ and\ \citenamefont {Marangos}}]{Fleischhauer2005}%
  \BibitemOpen
  \bibfield  {author} {\bibinfo {author} {\bibfnamefont {M.}~\bibnamefont {Fleischhauer}}, \bibinfo {author} {\bibfnamefont {A.}~\bibnamefont {Imamoglu}},\ and\ \bibinfo {author} {\bibfnamefont {J.~P.}\ \bibnamefont {Marangos}},\ }\bibfield  {title} {\bibinfo {title} {Electromagnetically induced transparency: Optics in coherent media},\ }\href {https://doi.org/10.1103/RevModPhys.77.633} {\bibfield  {journal} {\bibinfo  {journal} {Rev. Mod. Phys.}\ }\textbf {\bibinfo {volume} {77}},\ \bibinfo {pages} {633} (\bibinfo {year} {2005})}\BibitemShut {NoStop}%
\bibitem [{\citenamefont {Firstenberg}\ \emph {et~al.}(2016)\citenamefont {Firstenberg}, \citenamefont {Adams},\ and\ \citenamefont {Hofferberth}}]{Firstenberg2016}%
  \BibitemOpen
  \bibfield  {author} {\bibinfo {author} {\bibfnamefont {O.}~\bibnamefont {Firstenberg}}, \bibinfo {author} {\bibfnamefont {C.~S.}\ \bibnamefont {Adams}},\ and\ \bibinfo {author} {\bibfnamefont {S.}~\bibnamefont {Hofferberth}},\ }\bibfield  {title} {\bibinfo {title} {Nonlinear quantum optics mediated by rydberg interactions},\ }\href {https://doi.org/10.1088/0953-4075/49/15/152003} {\bibfield  {journal} {\bibinfo  {journal} {Journal of Physics B: Atomic, Molecular and Optical Physics}\ }\textbf {\bibinfo {volume} {49}},\ \bibinfo {pages} {152003} (\bibinfo {year} {2016})}\BibitemShut {NoStop}%
\bibitem [{\citenamefont {Gea-Banacloche}\ \emph {et~al.}(1995)\citenamefont {Gea-Banacloche}, \citenamefont {Li}, \citenamefont {Jin},\ and\ \citenamefont {Xiao}}]{GeaBanacloche1995}%
  \BibitemOpen
  \bibfield  {author} {\bibinfo {author} {\bibfnamefont {J.}~\bibnamefont {Gea-Banacloche}}, \bibinfo {author} {\bibfnamefont {Y.-q.}\ \bibnamefont {Li}}, \bibinfo {author} {\bibfnamefont {S.-z.}\ \bibnamefont {Jin}},\ and\ \bibinfo {author} {\bibfnamefont {M.}~\bibnamefont {Xiao}},\ }\bibfield  {title} {\bibinfo {title} {Electromagnetically induced transparency in ladder-type inhomogeneously broadened media: Theory and experiment},\ }\href {https://doi.org/10.1103/PhysRevA.51.576} {\bibfield  {journal} {\bibinfo  {journal} {Phys. Rev. A}\ }\textbf {\bibinfo {volume} {51}},\ \bibinfo {pages} {576} (\bibinfo {year} {1995})}\BibitemShut {NoStop}%
\bibitem [{\citenamefont {Xiao}\ \emph {et~al.}(1995)\citenamefont {Xiao}, \citenamefont {Li}, \citenamefont {Jin},\ and\ \citenamefont {Gea-Banacloche}}]{Banacloche952}%
  \BibitemOpen
  \bibfield  {author} {\bibinfo {author} {\bibfnamefont {M.}~\bibnamefont {Xiao}}, \bibinfo {author} {\bibfnamefont {Y.-q.}\ \bibnamefont {Li}}, \bibinfo {author} {\bibfnamefont {S.-z.}\ \bibnamefont {Jin}},\ and\ \bibinfo {author} {\bibfnamefont {J.}~\bibnamefont {Gea-Banacloche}},\ }\bibfield  {title} {\bibinfo {title} {Measurement of dispersive properties of electromagnetically induced transparency in rubidium atoms},\ }\href {https://doi.org/10.1103/PhysRevLett.74.666} {\bibfield  {journal} {\bibinfo  {journal} {Phys. Rev. Lett.}\ }\textbf {\bibinfo {volume} {74}},\ \bibinfo {pages} {666} (\bibinfo {year} {1995})}\BibitemShut {NoStop}%
\bibitem [{\citenamefont {Steck}(2019)}]{Steck2019}%
  \BibitemOpen
  \bibfield  {author} {\bibinfo {author} {\bibfnamefont {D.~A.}\ \bibnamefont {Steck}},\ }\href {http://steck.us/alkalidata} {\bibinfo {title} {Cesium d line data}} (\bibinfo {year} {2019})\BibitemShut {NoStop}%
\end{thebibliography}%



\pagebreak
\begin{widetext}
    
\begin{center}
\textbf{\large Supplemental Materials to All-optical radio-frequency phase detection for Rydberg atom sensors using oscillatory dynamics} \vspace*{10pt} \newline 
Matthias Schmidt,$^{1,2}$ Stephanie M. Bohaichuk,$^{1}$ Vijin Venu,$^{1}$ Ruoxi Wang,$^{1}$ Harald K\"ubler$^{1,2}$ and James P. Shaffer$^{1}$\\ \vspace*{5pt}
$^1$\textit{Quantum Valley Ideas Laboratories, 485 Wes Graham Way, Waterloo, ON N2L 0A7, Canada} \\
$^2$\textit{5. Physikalisches Institut, Universität Stuttgart, Pfaffenwaldring 57, 70569 Stuttgart, Germany} \\
(Dated: \today)
\end{center}
\end{widetext}

\setcounter{equation}{0}
\setcounter{figure}{0}
\setcounter{table}{0}
\setcounter{page}{1}
\makeatletter
\renewcommand{\theequation}{S\arabic{equation}}
\renewcommand{\thefigure}{S\arabic{figure}}

\section{Methods}\label{AppSec:Methods}
The optical response of the atom is calculated by solving the Lindblad master equation using the density matrix formalism. Since we are primarily concerned with the scattering of light by the probe laser beam, the $\rho_{12}$ matrix element is of central focus. $\rho_{12}$ can be used to calculate the expectation value of the induced transition dipole moment with which the probe laser interacts with the atoms, $\langle\hat{\mu}_\mathrm{12}\rangle = \mu_\mathrm{12} \rho_{21}+\mu_\mathrm{12}^\star\rho_{12}$. $\hat{\mu}_\mathrm{12}$ is the dipole operator for the probe laser transition and the dipole moment, $\mu_\mathrm{12} = \bra{1}\hat{\mu}_\mathrm{12}\ket{2}$. To determine $\rho_{12}$, the Lindblad master equation, 
\begin{equation}
\dot{\rho}=\frac{i}{\hbar} [\cal{H}, \rho]-\cal{L},
\label{AppEq:Theory_DensityMatrixEquation}
\end{equation}
is solved numerically. $\hbar$ is the reduced Planck constant.
The Hamiltonian, $\cal{H}$, describes the coupling of the laser light fields and the RF E-field to the atoms. Without loss of generality we choose a reference frame, where all the phase contributions are manifested on the RF transition~\cite{Morigi02,Morigi07PhaseVapor}. In this reference frame $\cal{H}$ is given by
\begin{align}
    {\cal{H}} &= \hbar \sum_{i=2}^{5} \Delta_i \ket{i}\bra{i} \nonumber\\
    &+  \frac{\hbar}{2} \big( \Omega_{12}\ket{1}\bra{2} + \Omega_{23}\ket{2}\bra{3} +\Omega_{34}\ket{3}\bra{4} \nonumber\\ &+ \Omega_\mathrm{RF}\,\mathrm{e}^{i\phi}\ket{4}\bra{5}+\Omega_{25}\ket{2}\bra{5} + \, \mathrm{h.c.}\big),
    \label{AppEq:Hamilton}
\end{align}
with $\Delta_2 = \Delta_{12}$, $\Delta_3 = \Delta_{12}+\Delta_{23}$, $\Delta_4 = \Delta_{12}+\Delta_{23}+\Delta_{34}$ and $\Delta_5 = \Delta_{12}+\Delta_{25}$ being the multi-photon detunings of the respective transition.
The Rabi frequencies of the individual driving fields are given by $\Omega_{ij} = \mu_{ij}E_{ij}/\hbar$, where $ij$ label respective transition. We use "RF" as a label instead of "45" to emphasize the transition of interest, i.e. $\Omega_{45}:=\Omega_\mathrm{RF}$. The dipole matrix element for each transition is denoted $\mu_{ij}$. $E_{ij}$ denotes the peak amplitude of each electric field. The probe laser transition drives the D1 transition of cesium and the excitation wavelength of the respective transition is shown in  Fig.~1(b). The overall phase, $\phi$, in the rotated basis reads
\begin{align}
    \phi &= \Delta_\mathrm{eff}\cdot t +  k_\mathrm{eff}\cdot z + \varphi_\mathrm{eff} \nonumber \\
    &= ( \Delta_{23}+\Delta_{34} + \Delta_\mathrm{RF}-\Delta_{25})\cdot t \nonumber \\ 
    &+ v(k_{23}+k_{34}+k_\mathrm{RF}-k_{25})\cdot t \nonumber \\
    &+ (k_{23}+k_{34}+k_\mathrm{RF}-k_{25}) \cdot z \nonumber \\ 
    &+ (\varphi_{23}+\varphi_{34}+\varphi_\mathrm{RF}-\varphi_{25}). 
    \label{AppEq:Phase}
\end{align}
It comprises the time-dependent term $\Delta_\mathrm{eff}\cdot t$, the spatial-dependent term $k_\mathrm{eff}\cdot z$, and the relative phase term $\varphi_\mathrm{eff}$. We assume that all lasers are locked to the same phase and the phase term simplifies to $\varphi_\mathrm{eff}=\varphi_\mathrm{RF}$. In the co-linear geometry proposed in Fig.~1(a) the excitation loop has a perfect $k$-vector compensation, $k_\mathrm{eff}=0$, and the system is spatially independent. Practically, the angle of arrival of the RF field compared to the optical axis should kept small so that the projection of the wave vector of the RF wave compensates $k_\mathrm{eff}$ so that the spatial periodicity, $1/k_\mathrm{eff}\gg L$, where $L$ is the length of the atomic medium, i.e. cell length. This geometrical constraint is crucial as the signal over a spatial period averages out.
The time-dependent term in \EqnRef{AppEq:Phase} consists of the laser detuning and the Doppler shifts of the atom. 
For the calculations, we assume that the RF wave as well as all lasers, except the blue laser, are on resonance, which leads to $\Delta_\mathrm{eff} = \Delta_{25}\cdot t$. The sign can be chosen freely under the described conditions. The other fields in the loop, or combinations thereof, can be used to induce oscillations of the interferometer by detuning them. The time dependent Doppler detuning term in the phase expression also depends on $k_\mathrm{eff}$. The $k_\mathrm{eff}$ dependence implies that for $k_\mathrm{eff}\neq0$, the atoms belonging to different velocity classes have different Doppler shifts, so their complex phase oscillates at different frequencies. 
We want to emphasize, that the negative sign in front of $\Delta_{25}$, $\varphi_{25}$ and $k_{25}$, corresponding to the blue laser, comes from the energetic position of the upper Rydberg state, which, in this example, is the highest energy state. The RF wave and the blue laser build a $\Lambda$-type subsystem in the excitation scheme, where the detuning picks up a negative sign~\cite{Fleischhauer2005}. 
For an energetically lower lying second Rydberg state $\ket{5}$, the RF transition would get a negative sign in \EqnRef{AppEq:Phase} and the system would be symmetric. 
\begin{figure}
\includegraphics[width = \columnwidth]{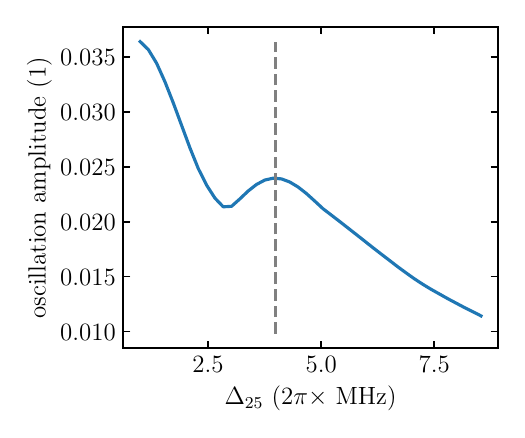}
\caption{Amplitude of the oscillatory dynamics of $I_{12}(t)$. The highest amplitude is reached for small $\Delta_{25}$ close to the atomic resonance. At $\Delta=\SIangfreq{3}{\mega \hertz }$, the amplitude has a minimum followed by a maximum at $\SIangfreq{4}{\mega \hertz}$. For larger $\Delta_{25}$, the amplitude continuously decreases. The dashed gray line corresponds to the chosen $\Delta_{25}=\SIangfreq{4}{\mega \hertz}$ for the calculations presented in the main text of this work. $\Omega_\mathrm{12}=\SIangfreq{100}{\kilo \hertz}$ and $\Omega_{23}=\Omega_{34}=\Omega_{25} = \SIangfreq{3}{\mega \hertz}$.}
\label{AppFig:DetuningAmp}
\end{figure}
For the calculations presented in the main text, we choose $\Delta_{25} = \SIangfreq{4}{\mega \hertz}$. This is based on the analysis in \FigRef{AppFig:DetuningAmp}, where the oscillation amplitude of $I_{12}(t)$ is shown as a function of $\Delta_{25}$. The oscillation amplitude reaches is maximum close to the atomic resonance at small $\Delta_{25}$. For large $\Delta_{25}$, the oscillation amplitude continuously decreases and eventually vanishes. In between, for $\Delta_{25}= \SIangfreq{3}{\mega \hertz}$, the amplitude has a minimum and shows a local maximum at $\SIangfreq{4}{\mega \hertz}$. Our proposed phase-sensitive RF detection schemes are based on demodulation techniques of $I_{12}(t)$. To achieve a successful and robust demodulation, the oscillation period of $I_{12}(t)$ has to be smaller than the target RF pulses, i.e. multiple oscillations have to be in one RF pulse. The RF pulse duration in the main text of the paper is $\SI{1}{\micro \second}$. Therefore, we chose the maximum amplitude at $\Delta_{25}=\SIangfreq{4}{\mega \hertz}$ for the calculations. 

The Lindblad operator, $\cal{L}$, describes the decay and dephasing of the populations and coherences~\cite{Kuebler2018, Bohaichuk2022, Schmidt2024}. In our numerical calculations, we use $\gamma_{2} = \SIangfreq{5}{\mega \hertz}$ for the decay rate of the $6P_{1/2}$ state and $\gamma_3 = \SIangfreq{1}{\mega \hertz}$ for the decay rate of the $9S_{1/2}$ state. For both Rydberg states, we assume total decay rates of $\gamma_4 = \gamma_5 = \SIangfreq{20}{\kilo \hertz}$. The values for $\gamma_4$ and $\gamma_5$ are used as representative decay rates for a Rydberg state. We include a transit time broadening of $\SIangfreq{200}{\kilo \hertz}$. Collisional dephasing and laser dephasing are neglected in our calculations to simplify the picture, but can be introduced in the Lindblad operator in a straightforward manner.

In a finite temperature vapor, the atoms move with a distribution of different velocities. 
The atoms experience Doppler shifts proportional to their velocity, $v$, that cannot be neglected in most Rydberg atom vapor cell electrometers. The Doppler shifts appear in the Hamiltonian, \EqnRef{AppEq:Hamilton}, and hence the density matrix equations, \EqnRef{AppEq:Theory_DensityMatrixEquation}, as velocity dependent detunings,
\begin{equation}
\Delta_{ij} =\vec{k}_{ij} \cdot \vec{v},
\label{AppEq:Thermal_DopplerShift}
\end{equation}
where $|\vec{k}_{ij}|= 2\pi/\lambda_{ij}$ denotes the magnitude of the wave vector of the laser whose wavelength is $\lambda_{ij}$. Since the lasers are co-linear in our geometry, \EqnRef{AppEq:Thermal_DopplerShift} is a scalar, where the sign of $k_{ij}$ denotes the propagation direction of the lasers relative to one another.

The atomic response, $\epsilon$, can be calculated as
\begin{equation}
\epsilon =\eta+ i \alpha = \frac{3\pi  n \lambdabar^3_\mathrm{12} }{2} k_\mathrm{12} \frac{\gamma_2}{\Omega_\mathrm{12}} \rho_{12}  = A k_\mathrm{12} \frac{\gamma_2}{\Omega_\mathrm{12}} \rho_{12}.
\label{AppEq:AbsorptionCoefficientDefinition}
\end{equation}
$\epsilon$ depends on the laser fields and the RF field, including their characteristics like amplitude, detuning and phase. $\epsilon$ describes the refractive index, $\eta$, and absorption, $\alpha$. $\epsilon$ is directly proportional to the transition dipole moment of the probe transition. 
The density of atoms, $n$, can be calculated using a temperature dependent vapor pressure model~\cite{Steck2019}. The reduced wavelength of a photon emitted on the probe transition is $\lambdabar_\mathrm{12} = \lambda_\mathrm{12}/2\pi$. The scattering of photons on the probe transition is modified by the incident RF electromagnetic waves interacting with the atoms through $\rho_\mathrm{12}$. $A$ is proportional to the number of atoms in a probe transition wavelength mode volume. 

The atomic response of a thermal atomic vapor is calculated by integrating over all atomic velocity classes weighted by the Maxwell-Boltzmann distribution,  at a given temperature, $T$,
\begin{equation}
\epsilon = A k_\mathrm{12} \frac{\gamma_2}{\Omega_\mathrm{12}} \int_{-\infty}^{\infty} P(v) \rho_{12}(v) \, \mathrm{d}v.
\label{AppEq:AtomicResponseDefinition}
\end{equation}
The Maxwell-Boltzmann velocity distribution of the gas is $P(v)= 1/\sqrt{\pi}\bar{v} \,\exp\left(-v^2/\bar{v}^2\right)$ with the atomic mass, $m$, the Boltzmann constant, $k_\mathrm{B}$, and the average speed of the thermal cloud, $\bar{v}=\sqrt{2k_\mathrm{B} T/m}$. We use room temperature, $T=\SI{300}{\kelvin}$, for the calculations, unless otherwise noted. 

The transmitted intensity of the probe laser beam through the vapor cell is described by the Lambert-Beer law
\begin{equation}
I_{12}(t) = I_0\, \mathrm{e}^{-\alpha(t) L}.
\label{AppEq:Intensity}
\end{equation}
where the initial probe laser intensity, $I_0= c|E_{12}|^2/8\pi$. The length of the vapor cell, i.e. distance traveled through the atomic medium, is $L= \SI{3}{\centi \meter}$. 

\section{Weak Probe approximation}\label{AppSec:WeakProbeApprox}
\begin{figure*}
\includegraphics[width=.19\textwidth]{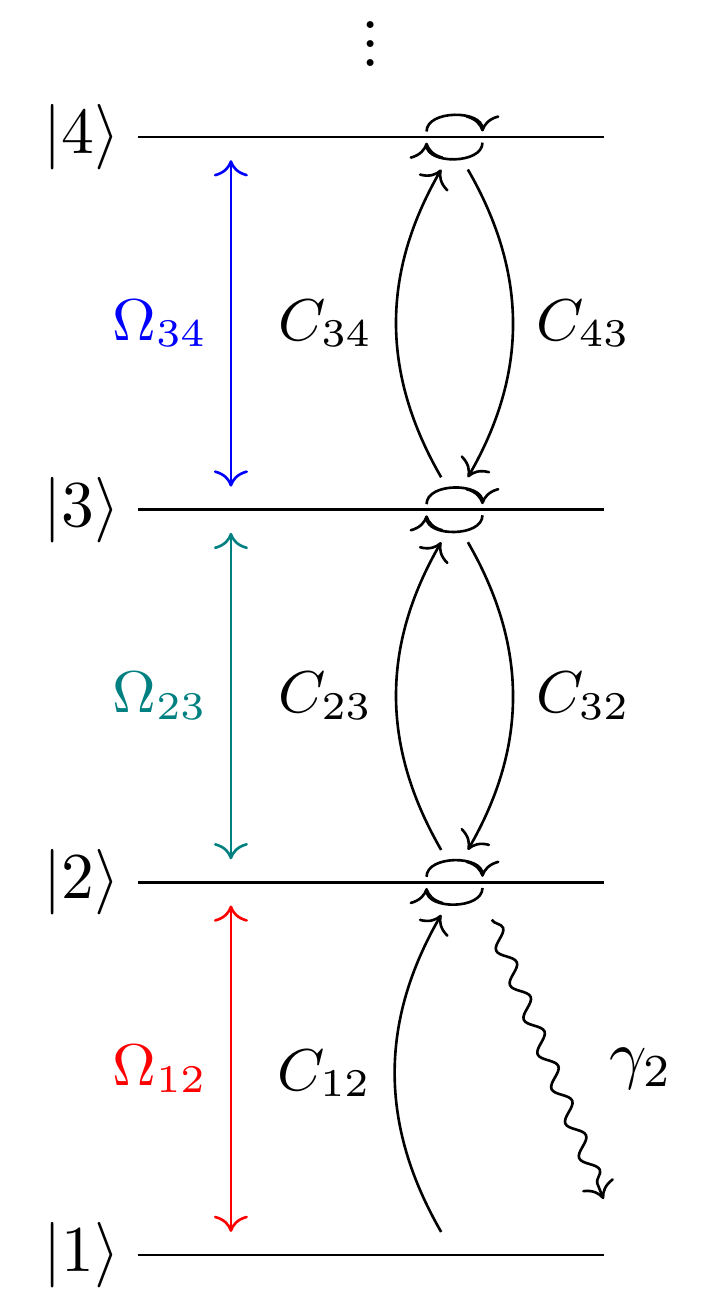}
\includegraphics[width=.44\textwidth]{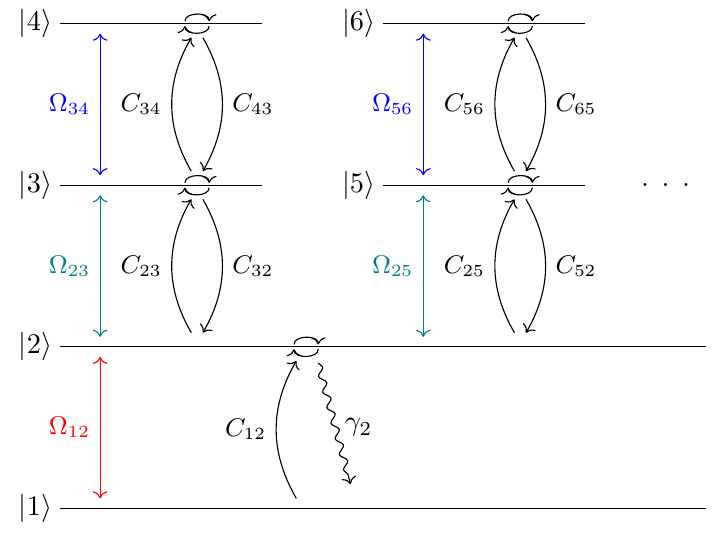} 
\includegraphics[width=.33\textwidth]{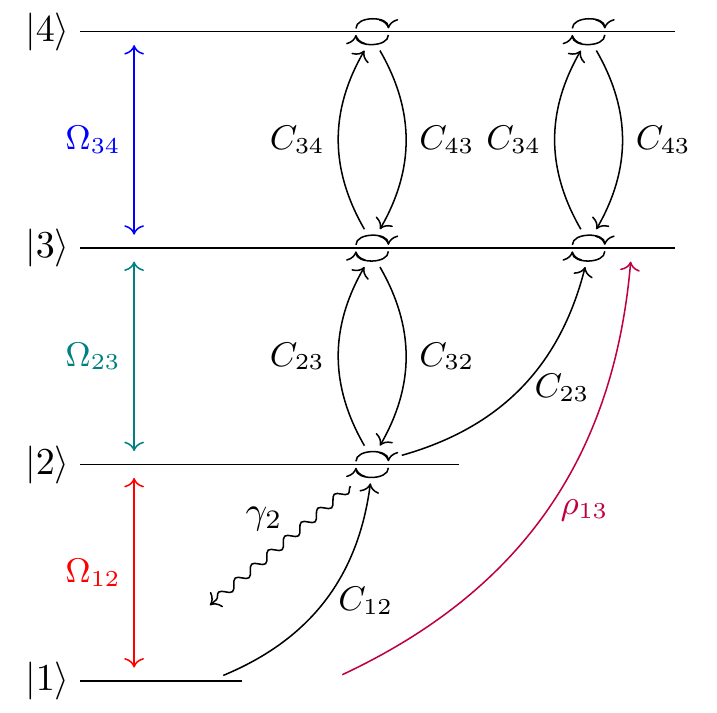}
\put(-495,175){(a)} \put(-395,175){(b)}\put(-180,175){(c)}
\put(-470,175){Ladder scheme} \put(-305,175){Y-type scheme}\put(-140,175){Multi-photon coherence transfer}
\caption{Illustrations of the basic interference processes for various excitation schemes. Only the lowest order processes of the coupling coherences are considered here. The weak probe approximation is assumed. (a) Ladder excitation scheme. The coherences lead to constructive and destructive interference with the probe absorption, i.e. EIT and EIA. (b) Y-type scheme. For excitation schemes where a branching is present, the individual interference pathways are superposed. (c) Multi-photon coherence transfer. The coupling fields establish a coherence between the levels not directly coupled to each other.}
\label{AppFig:Basic_Interference_Processes}
\end{figure*}
In the weak probe approximation, $\Omega_\mathrm{12}$ is considered to be significantly smaller than the effective  Rabi frequencies of the loop, as well as the decay rate of the excited state, $\gamma_2$. Effective Rabi frequencies means the geometrical mean of the involved Rabi frequencies in the multi-photon interference processes investigated here.
Consequently, $\Omega_\mathrm{RF}$ can be of the same magnitude or even smaller than the probe Rabi frequency and the weak probe approximation is still valid for strong optical coupling fields.
The dominant decay process of the excited atoms is therefore the spontaneous decay of the upper state of the probed transition. Under this assumption, terms of higher order in $\Omega_{12}$ are neglected~\cite{Fleischhauer2005,Schmidt2024,Firstenberg2016, GeaBanacloche1995, Banacloche952}.
To derive Eq.~(4), we considered the underlying multi-photon interference processes in the closed-loop excitation scheme. We define the coherent coupling of two atomic states via an electromagnetic field as $C_{jk}:=i\Omega_{jk}/(\gamma_k+2i\Delta_{k})$. In the definition of the coherences, the multi-photon detunings $\Delta_{k}$ are considered.
We interpret $C_{jk}$ for $k>j$, and correspondingly the energy of the state ${\cal E }_k>{\cal E }_j$, as a coherent excitation and $C_{jk}$ with $j>k$ as stimulated de-excitation, ${\cal E }_j>{\cal E }_k$. Note that in our excitation scheme the labelling of the states is ordered according to their energy, i.e. ${\cal E }_1$ has the lowest energy and ${\cal E }_5$ the highest.
Note that the coherent de-excitation is not the complex conjugate of the coherent excitation.
The coherent de-excitation differs in two ways from the coherent excitation. Firstly, $\Delta_k$ and $\gamma_k$ are different for the respective process as the properties of the $k$-state determine the coherence. Secondly, while the magnitude of the Rabi frequencies for the two processes are the same, $|\Omega_{jk}|=|\Omega_{kj}|$, the Rabi frequency of the de-excitation follows $\Omega_{jk}=\Omega_{kj}^\star$, in analogy to the Hermitian conjugated term in \EqnRef{AppEq:Hamilton}.

The coherences spectrally take the shape of a complex Lorentzian. Depending on the strength of the driving field, $\Omega_{ij}$, the detunings and the decay rates, the excitation pathways, built up by the coherences experience different phase shifts that lead to constructive or destructive interference of the amplitude of the atomic response, similar to EIT. 
Here, we extend typical three-level EIT, based on interference between two coherences, to multi-photon EIT by considering the lowest order interference processes, i.e. pathways through the excitation scheme, of the coherences required to construct the atomic response for the excitation scheme presented in Fig.~1(b). 
To understand the underlying interference processes, we first consider the fundamental building blocks of the interferences.
The basic building blocks needed to describe the closed-loop system are illustrated in \FigRef{AppFig:Basic_Interference_Processes}. In \SubFigRef{AppFig:Basic_Interference_Processes}{a} a four level ladder excitation scheme is sketched on the left. On the right, the corresponding lowest order coherence processes are shown. The coherences build on each other as the runs of the ladder are traversed. The atomic response of such an excitation scheme can be expressed as a continued fraction~\cite{Schmidt2024},
\begin{equation}
    \rho_{12}= \dfrac{ C_\mathrm{12} }{1+ \dfrac{ C_\mathrm{23}C_\mathrm{32} }{ 1+\dfrac{C_\mathrm{34}C_\mathrm{43} }{1+\ddots} } }.
    \label{AppEq:Ladder}
\end{equation}
The continued fraction consists of a fraction for each pair of levels that are coupled to the probe transition, e.g. three-photon excitation schemes with two coupling laser transitions have two fractions in the denominator~\cite{Bohaichuk2023threephoton, Venu2025}, indicated by the dots in \EqnRef{AppEq:Ladder} and \SubFigRef{AppFig:Basic_Interference_Processes}{a}. For three levels, the continued fraction describes a typical EIT scheme with a single fraction in the denominator. A fourth level leads to enhanced absorption and is referred to as electromagnetically induced absorption. Each added level alternates between enhanced transmission and enhanced absorption, corresponding to alternating destructive and constructive interference, respectively. The arrows that create a circle on each excited state illustrate the "1+" terms on each continued fraction, corresponding to the interference of the state with itself.

The excitation scheme presented in \SubFigRef{AppFig:Basic_Interference_Processes}{b} shows a branching of transitions coupled to the upper state of the probe transition, i.e. two transitions are driven from $\ket{2}$, $\ket{2}\rightarrow\ket{3}$ and $\ket{2}\rightarrow\ket{5}$, by different Rabi frequencies,$\Omega_{23}$ and $\Omega_{25}$, respectively. The individual branches are coupled to another level by an additional driving field.
In the closed-loop system in Fig.~1(b), the branching occurs on the upper state of the probe transition.
The corresponding atomic response is given by
\begin{align}
    \rho_{12}=\dfrac{ C_\mathrm{12} }{ 1+\dfrac{C_\mathrm{23}C_\mathrm{32} }{ 1+ C_\mathrm{34}C_\mathrm{43} } + \dfrac{C_\mathrm{25}C_\mathrm{52} }{ 1+ C_\mathrm{56} C_\mathrm{65} } + \dots }.
    \label{AppEq:Branching}
\end{align}
Each branch in \SubFigRef{AppFig:Basic_Interference_Processes}{b} is a four level ladder excitation process similar to \SubFigRef{AppFig:Basic_Interference_Processes}{a}. The corresponding continued fraction form appears similarly in \EqnRef{AppEq:Branching}.
The individual branches add in the denominator of \EqnRef{AppEq:Branching}, implying that the two ladder excitation branches are not independent of each other. The excitation pathways interfere on the upper level of the probe transition. Consequently, one branch affects the other branch similar to a beam splitting interferometer. In other words, if $\Omega_{56}$ increases, the induced absorption from $\Omega_{34}$ on the probe transition decreases.
The dots in \SubFigRef{AppFig:Basic_Interference_Processes}{b} illustrate that the superposition of branches is not limited, similar to the continued fraction for the ladder excitation schemes discussed in (a).

The closed-loop section of the excitation scheme in Fig.~1(b) is a multi-photon transfer process. The multi-photon transfer process describes the established coherence between two states that are established via multiple transitions.
\SubFigRefCap{AppFig:Basic_Interference_Processes}{c} illustrates a multi-photon coherence transfer, where the coherence between $\ket{1}$ and $\ket{3}$, i.e. $\rho_{13}$, is analyzed in a four level system. The transfer from the ground state, $\ket{1}$, to $\ket{3}$ occurs via $C_{12}$ and $C_{23}$. Both coherences interfere with all higher lying transitions. The multi-photon coherence transfer is expressed as
    \begin{align}
    \rho_{13} &= \dfrac{ C_\mathrm{12} }{1+\dfrac{ C_\mathrm{23}C_\mathrm{32} }{ 1+C_\mathrm{34}C_\mathrm{43} } } \cdot \dfrac{ C_\mathrm{23} }{ 1+ C_\mathrm{34}C_\mathrm{43} } \nonumber \\ &= \frac{ C_{12}C_{23} }{ 1+C_{23}C_{32}+C_{34}C_{43} }.
\end{align}
The two-photon coherence transfer illustrated here is given by a multiplication of the two coherences, $C_{12}$ and $C_{23}$ and their respective interference pathway with higher lying transitions. Each of the transfer coherences carries the interference information of the ladder excitation with it, leading to the continued fraction. It is important to note that the transfer coherences occur linearly, whereas all terms in the interference processes presented so far appear quadratic, i.e. the absolute value, in their respective Rabi frequencies. The linearity of the transfer excitation is especially important for the closed-loop processes in Fig.~1(b) as it carries the phase information.

\begin{figure*}
    \includegraphics[width= \textwidth]{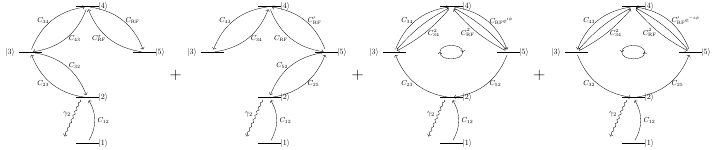}
    \put(-495,105){(a)}\put(-375,105){(b)}\put(-250,105){(c)}\put(-120,105){(d)}
    \caption{Illustrations of the underlying multi-photon interference processes in the closed-loop system. All processes are linear in $C_{12}$ and predominantly decay via $\gamma_2$. In the weak probe approximation, we consider only the lowest order coherence processes. The first two processes, (a) and (b), correspond to the left- and right-handed ladder absorption processes. The third and fourth processes correspond to the counterclockwise and clockwise interference loops that are explicitly phase dependent, respectively in (d) and (c). The states in the illustrations do not represent the energetic ordering of the states accurately but are drawn to illustrate the topology.}
    \label{AppFig:WeakProbe_Processes}
\end{figure*}
In the case of the closed-loop excitation scheme, we find four different processes that need to be considered to derive the analytical expression from Eq.~(4).
The processes are illustrated in \FigRef{AppFig:WeakProbe_Processes}. The state labels in the illustrations do not represent the ordering of the energies of the levels accurately because we emphasize the topology of each process. $\ket{5}$ is considered to have a higher excitation energy than the other levels, in particular, compared to the second Rydberg state. An energy ordered diagram of the excitation scheme is found in Fig.~1(a). In \SubFigRef{AppFig:WeakProbe_Processes}{a} and (b), the first two diagrams show the absorptive response of the atoms, and are described by a continued fraction found in Ref.~\cite{Schmidt2024}.
In \SubFigRef{AppFig:WeakProbe_Processes}{a}, the excitation follows the left-handed path through the excitation scheme, whereas in (b), the excitation follows the right-hand ladder. Both ladder excitation terms depend quadratically on the respective Rabi frequencies, in analogy to the discussion of \SubFigRef{AppFig:Basic_Interference_Processes}{a}. The phase contributions cancel out and the pair of terms are not phase dependent.
It is noteworthy, that the RF transition induces enhanced transmission in (a) and increases absorption in (b), similarly the $\SI{2.2}{\micro \meter}$ transition shows the same behavior. The relative magnitudes of the Rabi frequencies on the loop transitions decide which pathway dominates and therefore whether the offset of the oscillatory dynamics shifts towards increased or decreased probe laser absorption relative to the steady-state transmission when $\Omega_\mathrm{RF}$ is absent.
In \SubFigRef{AppFig:WeakProbe_Processes}{c} and (d), the excitation processes illustrated are the clockwise and counter-clockwise excitation loops. The loops can be seen as four-photon coherence transfer from $\ket{2}$ to $\ket{2}$ and can occur around a counterclockwise or clockwise path. Similarly to the discussion in \SubFigRef{AppFig:Basic_Interference_Processes}{c}, the coherences in the closed-loop processes appear linearly. Therefore, the closed-loop terms have an intrinsic phase dependency. Depending on the direction of rotation, the phase term appears as $\mathrm{e}^{i\phi}$ or $\mathrm{e}^{-i\phi}$ for the clockwise and counterclockwise processes, respectively. The additional denominator of the closed-loop term in Eq.~(4) comes from the additional of the loops on the $\SI{2.2}{\micro \meter}$- and RF-transition, reminiscent of the discussion on \SubFigRef{AppFig:Basic_Interference_Processes}{c}.
All four of the processes contributing to $\rho_{12}$ are needed to explain the atomic response accurately. The branching of the the four processes occurs on the upper state of the probe transition. The atomic response on the probe transition is then expressed as
\begin{widetext}
\begin{small}
\begin{align}
\rho_{12} &=\dfrac{C_\mathrm{12}}{1+\dfrac{C_\mathrm{23}C_\mathrm{32}}{1+\dfrac{C_\mathrm{34}C_\mathrm{43}}{1+ C_\mathrm{RF} C_\mathrm{RF} }} + \dfrac{C_\mathrm{25}C_\mathrm{52}}{1+\dfrac{C_\mathrm{RF}C_\mathrm{RF}}{1+C_\mathrm{34}C_\mathrm{43}}} - \dfrac{ C_\mathrm{23} }{1+\dfrac{C_\mathrm{34}C_\mathrm{43} }{1+C_\mathrm{RF}C_\mathrm{RF}}}  \dfrac{ C_\mathrm{34} }{1+C_\mathrm{RF} C^\prime_\mathrm{RF}}   C_\mathrm{RF}\mathrm{e}^{i\phi}  C_\mathrm{52}- \dfrac{ C_\mathrm{25} }{1+\dfrac{C_\mathrm{RF}C^\prime_\mathrm{RF}}{1+C_\mathrm{43}C_\mathrm{34}}}  \dfrac{ C^\prime_\mathrm{RF}\mathrm{e}^{-i\phi} }{1+C_\mathrm{43}C_\mathrm{34} } C_\mathrm{34} C_\mathrm{32}} \nonumber\\ 
    &=\dfrac{C_\mathrm{12}}{1+\dfrac{C_\mathrm{23}C_\mathrm{32} }{1+\dfrac{C_\mathrm{34}C_\mathrm{43} }{1+ C_\mathrm{RF}C^\prime_\mathrm{RF} }} + \dfrac{C_\mathrm{25}C_\mathrm{52}}{1+\dfrac{C_\mathrm{RF}C^\prime_\mathrm{RF}}{1+ C_\mathrm{34}C_\mathrm{43} }} - \dfrac{2C_\mathrm{23}C_\mathrm{34}C_\mathrm{RF}C_\mathrm{52} \mathrm{cos}(\phi) }{ 1+ C_\mathrm{34}C_\mathrm{43} + C_\mathrm{RF}C^\prime_\mathrm{RF}} }.
    \label{AppEq:WeakProbeExpression}
\end{align}
\end{small}
\end{widetext}
The counterclockwise and clockwise terms simplify to the $\cos(\phi)$-term, that describes the phase dependence of the system in lowest order. The system shows a $2\pi$-periodicity. $C^\prime_\mathrm{RF}$ corresponds to the de-excitation on the RF transition. We chose $C^\prime_\mathrm{RF}$ explicitly instead of $C^\star_\mathrm{RF}$ to emphasize that the coherent de-excitation is not the complex conjugate of the coherent excitation.

\section{Bandwidth analysis}\label{AppSec:Bandwidth} 
The bandwidth of the response is determined by the transient time scale of the atomic response upon the arrival of the RF pulse. 
\begin{figure}
\includegraphics[width = \columnwidth]{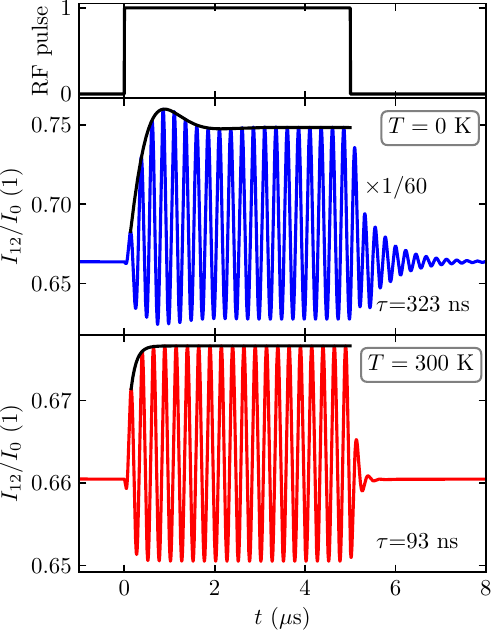}
\put(-250,310){(a)}\put(-250,260){(b)}\put(-250,140){(c)}
\caption{(a) Incoming $\SI{5}{\micro \second}$ RF pulse with $\Omega_\mathrm{RF}=\SIangfreq{0.5}{\mega \hertz}$. (b), (c) Comparison of calculations at zero-temperature, $\SI{0}{\kelvin}$, and a thermal ensemble of atoms at $T=\SI{300}{\kelvin }$. in the upper and lower panel, respectively. The transmitted intensity of the probe laser as a function of time and is described as a $\SI{5}{\micro \second}$ RF pulse with the same phase for $T=0\,$K (b) and $T=300\,$K (c). The transient time scale for the thermal ensemble is $\approx 3.5\times$ faster compared to zero-temperature. Similar timescale difference is observed for the post pulse decay of the signal. The atomic response in the zero temperature case is $\approx60\times$ stronger than in the thermal limit and is rescaled for better comparison. $\Delta_{25}/2\pi= \SI{4}{\mega \hertz}$, $\Omega_\mathrm{12}=\SIangfreq{100}{\kilo \hertz}$ and $\Omega_{23}=\Omega_{34}=\Omega_{25} = \SIangfreq{3}{\mega \hertz}$.}
\label{Fig:Bandwidth}
\end{figure}
To estimate the transient time scale, we fit 
\begin{equation}
    f(t) = a \mathrm{e}^{-t/\tau}\cos(\omega t+\beta) + d
    \label{AppEq:Fit}
\end{equation}
to the envelope of the probe laser intensity oscillations. $\tau$ corresponds to the $1/\mathrm{e}$-timescale of the transients. The fit parameters $\alpha$, $\omega$ and $\beta$ in \EqnRef{AppEq:Fit} denote the amplitude, frequency and offset phase of the slowly varying oscillation of the envelope, respectively. We focus on the effects of dephasing on the bandwidth. In our system, the largest dephasing present is the thermal motion of the atoms. We compare the calculation of a thermal vapor to a gas of zero temperature atoms. The results are shown in \SubFigRef{Fig:Bandwidth}{b} and (c). The zero temperature limit calculation shows intensity oscillations at $\Delta_{25}= \SIangfreq{4}{\mega \hertz}$. 
In the beginning of the pulse, the transients lead to an overshoot of the response before the steady-state  of the oscillations are reached. The overshoot motivates the oscillatory term in the fit function in \EqnRef{AppEq:Fit}. The oscillatory dynamics continues until the RF pulse ends. Afterwards, the system shows a decay time that lasts until the steady-state absorption is reached.
Note that the oscillations are shifted strongly towards enhanced transmission compared to the steady-state absorption when  $\Omega_\mathrm{RF}$ is absent. In the zero temperature limit, the same effect is strongly enhanced due to the lack of dephasing due to Doppler shifts. 
The black curves in the respective subfigures correspond to the fits using \EqnRef{AppEq:Fit}. The time scales for the decay in the two cases are $\tau = \SI{323}{\nano \second}$ for the zero temperature limit and $\tau = \SI{95}{\nano \second}$ for the thermal vapor. The bandwidth is around $3.5\times$ smaller than for the zero temperature case.  
The overall signal strength in the zero temperature limit is approximately $60\times$ larger than in the thermal vapor which is similar to the findings in~\cite{Schmidt2024}. The curve of the probe intensity in \SubFigRef{Fig:Bandwidth}{b} is scaled by that factor for better visual comparison.

\end{document}